\documentclass{aa}  
\usepackage{graphicx}
\usepackage{textcomp}
\usepackage{txfonts}
\usepackage{cuted}
\usepackage{answers}
\usepackage{natbib}
\usepackage{amsmath}
\newcommand\SortNoop[1]{}
\begin{document} 

   \title{Distribution of shape elongations of main belt asteroids derived from Pan-STARRS1 photometry}

   \author{H. Cibulkov\'a \inst{1} \and H. Nortunen \inst{2} \and J. \v{D}urech \inst{1} \and M. Kaasalainen \inst{2} \and P. Vere\v{s} \inst{3} \and R. Jedicke \inst{4} \and R. J. Wainscoat \inst{4} \and M. Mommert \inst{5} \and D.~E.~Trilling \inst{5} \and E. Schunov\'{a}-Lilly \inst{4} \and E. A. Magnier \inst{4} \and C. Waters \inst{4} \and H. Flewelling \inst{4}
          }
       
   \institute{Institute of Astronomy, Faculty of Mathematics and Physics, Charles University,
              V Hole\v{s}ovi\v{c}k\'ach 2, 180 00 Prague 8\\
              \email{cibulkova@sirrah.troja.mff.cuni.cz}
         \and Department of Mathematics, Tampere University of Technology, PO Box 553, 33101 Tampere, Finland
         \and {Harvard-Smithsonian Center for Astrophysics, 60 Garden St, Cambridge, MA 02138, USA}
         \and {Institute for Astronomy University of Hawaii, 2680 Woodlawn Dr, Honolulu, HI, 96822, USA}
         \and{Department of Physics and Astronomy, PO Box 6010, Northern Arizona University, Flagstaff, AZ 86011, USA}
                     }

   \date{Received 07/2017; Accepted 10/09/2017}

 
  \abstract
   {A lot of photometric data is produced by surveys such as Pan-STARRS, LONEOS, WISE or Catalina. These data are a~rich source of information about the physical properties of asteroids. There are several possible approaches for utilizing these data. Lightcurve inversion is a typical method that works with individual asteroids. Our approach in this paper is statistical when we focused on large groups of asteroids like dynamical families and taxonomic classes, and the data were not sufficient for individual models.}
    {Our aim was to study the distributions of shape elongation $b/a$ and the spin axis latitude $\beta$ for various subpopulations of asteroids and to compare our results, based on Pan-STARRS1 survey, with statistics previously done using different photometric data (Lowell database, WISE data).}
    {We use the LEADER algorithm to compare the $b/a$ and $\beta$ distributions for different subpopulations of asteroids. The algorithm creates a cumulative distributive function (CDF) of observed brightness variations, and computes the $b/a$ and $\beta$ distributions using analytical basis functions that yield the observed CDF. A variant of LEADER is used to solve the joint distributions for synthetic populations to test the validity of the method.}
    {When comparing distributions of shape elongation for groups of asteroids with different diameters $D$, we found that there are no differences for $D<25\,$km. We also constructed distributions for asteroids with different rotation periods and revealed that the fastest rotators with $P = 0 - 4\,$h are more spheroidal than the population with $P = 4 - 8\,$h.}
    {}

   \maketitle

\section{Introduction}
The spin states (rotational periods and directions of the spin axes) and shapes of individual asteroids can be determined from photometric data by lighcurve inversion (Kaasalainen \& Lamberg \citeyear{2006InvPr..22..749K}, \v{D}urech et al. \citeyear{2015aste.book..183D} and references therein). For these methods, mainly dense photometric data are used, because they sample well the rotational period $P$. The preliminary estimate of $P$ can substantially reduce the computational time required for the determination of unique sidereal rotational period. Up to now, almost a thousand models have been derived using this method and most of them are stored in Database of Asteroids Models from Inversion Techniques (DAMIT; \v{D}urech et al. \citeyear{2010A&A...513A..46D}).

A different approach, suitable for photometric data sparse in time that are produced by all-sky surveys and consist typically of few measurements per night over $\sim10$ years, but not suitable for ordinary sparse lightcurve inversion (\v{D}urech et al. \citeyear{2005EM&P...97..179D}, \citeyear{2007IAUS..236..191D}), was described in \citet{2016A&A...596A..57C}. There, we used the mean brightness and its dispersion in individual apparitions to derive the ecliptical longitude and latitude of the spin axis and the shape elongation of asteroids from photometric data stored in the Lowell Observatory database (Bowell et al. \citeyear{2014M&PS...49...95B}). Even though the parameters could be determined for individual asteroids, the uncertainties are large and the results are supposed to be used in a statistical sense only. However, this model cannot be used for the photometric data from the Panoramic Survey Telescope \& Rapid Response System (Pan-STARRS), because there are not enough measurements covering long enough time intervals.

Another statistical study was done by \citet{2017A&A...601A.139N} using data from the WISE database\footnote{http://irsa.ipac.caltech.edu/Mission/wise.html}. These data also cannot be analyzed with the method from \citet{2016A&A...596A..57C}, however, \citet{2017A&A...601A.139N} developed a new model and described physical parameters for subpopulations of asteroids using distribution functions. This method is not meant to invert the shape and spin characteristics of individual lightcurves; the inversion works only on a population-scale, where we consider the shape and spin distributions of a large population. In \citet{2017A&A...601A.139N} as well as in this paper, we constructed cumulative distribution functions (CDFs) of the variation of brightness for selected groups of asteroids and studied the inverse problem. The parameters of the model are the shape elongation $b/a$ and the ecliptical latitude $\beta$ of the spin axis. The advantage of this method is that it can be used even if only few points and one apparition are available for an asteroid. A similar approach was used by \citet{2008Icar..196..135S} and \citet{2016MNRAS.459.2964M}.

While in \citet{2017A&A...601A.139N} we studied mainly the validity and accuracy of the method and practical applicability on astronomical databases, in this work, we applied the model on photometric data from Pan-STARRS1 and performed an analysis focusing on large subpopulations of asteroids using the Latitudes and Elongations of Asteroid Distributions Estimated Rapidly (LEADER) algorithm (Nortunen \& Kaasalainen, \citeyear{2017A&A..N}). The structure of this paper is as follows: In Sect. 2, we briefly describe the model; in Sect. 3, we describe the used data from Pan-STARRS1 sky survey; in Sect. 4, we test the accuracy of the determination of model parameters by simulations on synthetic data; in Sect. 5, we construct distributions of shape elongations $b/a$ and ecliptical latitudes $\beta$ of the spin axis for some subpopulations of asteroids and analyze the results and, in Sect. 6, we summarize the main results.

\section{Model}
In our model, we approximate the shape of an asteroid with a~simple, biaxial ellipsoid. We denote the semiaxes $a \ge b = c=1$, and we choose $b/a$ as the parameter that describes the shape elongation of an asteroid. We have $0 < b/a \le 1$, with a small $b/a$ presenting an elongated body, and $b/a = 1$ presenting a sphere. This shape approximation is very coarse, but with a high number of observations ($\propto 10^3$), it will portray statistical tendencies of a population accurately. For realistic shapes, the proportion of highly elongated values $b/a < 0.4$ is negligible, and for most shapes, $b/a > 0.5$. However, for completeness, we include all the values $0 < b/a \le 1$ in our grid; if the solved $b/a$ distribution contains an unusually high proportion of values below 0.4, it is usually an indicator of error in the solution, caused by noise and/or instabilities.

Our second parameter is the spin co-latitude $\beta$, defined as the ecliptic polar angle of the spin axis. The connection between $\beta$ and the aspect angle of the pole is explained in \citet{2017A&A...601A.139N}. In the ellipsoid model, the values of $\beta$ are fixed in the interval $[0, \pi/2]$. In other words, there is no way to distinguish whether the spin latitudes are above or below the ecliptic plane in our model. In our convention, $\beta = 0$ indicates that the spin direction is perpendicular to the ecliptic plane, while $\beta = \pi/2$ means the spin is in the ecliptic plane (it was this way in Nortunen et al. \citeyear{2017A&A...601A.139N}, but opposite in Cibulkov\'{a} et al. \citeyear{2016A&A...596A..57C}). We assume that most orbits are in the ecliptic plane.

Assuming we have the brightness intensities $L$ measured (with the data given by an asteroid database), we utilize the brightness variation $\eta$ as our observable:
\begin{equation}
\eta = \frac{\Delta (L^2)}{\langle L^2\rangle} .
\label{eq:eta}
\end{equation}
The squared intensities $L^2$ were used for the mean $\langle L^2\rangle$ and the variation $\Delta (L^2) := \sqrt{\langle(L^2-\langle L^2\rangle)^2\rangle}$ instead of the standard brightness $L$ in order to obtain more simple, closed-form formula for $\eta$. From \citet{2017A&A...601A.139N}, the amplitude $A$ can be directly computed from $\eta$:
\begin{equation}
A=\sqrt{1-\Big(\frac{1}{\sqrt{8}\eta}+\frac{1}{2}\Big)^{-1}} .
\label{eq:eta-A}
\end{equation}
Note that the amplitude $A$ is based on intensity here, not on magnitudes. With the amplitudes known, we can create their CDF, $C(A)$.

To solve the joint distribution for $b/a$ and $\beta$, we create a grid of bins $((b/a)_{\rm i}, \beta_{\rm j}) \in [0, 1] \, \times \, [0, \pi/2]$, where ${\rm i}=1$, $\ldots$, $k$ and ${\rm j}=1$, $\ldots$, $l$. Our goal is to determine the proportion of each bin. Now, the CDF can be written as a linear combination of other functions:
\begin{equation}
C(A) = \sum_{\rm i, j} w_{\rm ij} F_{\rm ij}(A) ,
\label{eq:series}
\end{equation}
where $F_{\rm ij}(A)$ are monotonously increasing basis functions derived by \citet{2017A&A...601A.139N}:
\begin{equation}
F_{\rm ij}(A)=\left\{\begin{array}{rc}
0, & A\le (b/a)_{\rm i}\\
\rule{0cm}{4.5ex}
\frac{\pi}{2}-\arccos\frac{\sqrt{A^2-(b/a)_{\rm i}^2}}{\sin\beta_{\rm j}\sqrt{1-(b/a)_{\rm i}^2}}\,,& (b/a)_{\rm i}<A<\mathcal{F}((b/a)_{\rm i}, \beta_{\rm j})\\
\rule{0cm}{4.5ex}
\frac{\pi}{2}, & A\ge\mathcal{F}((b/a)_{\rm i}, \beta_{\rm j}),
\end{array}\right.
\label{Fijeq}
\end{equation}
where $\mathcal{F}((b/a)_{\rm i}, \beta_{\rm j}) = \sqrt{\sin^2\beta_{\rm j}+(b/a)_{\rm i}^2\cos^2\beta_{\rm j}}$. Each basis function $F_{\rm ij}(A)$ describes the contribution made by objects in a given bin $((b/a)_{\rm i}, \beta_{\rm j})$ to the CDF $C(A)$. The weights $w_{\rm ij}$ are the occupation numbers of each bin $((b/a)_{\rm i}, \beta_{\rm j})$. We can write \eqref{eq:series} in an equivalent form,
\begin{equation}
Mw = C ,
\label{eq:lineq}
\end{equation}
where each column of the matrix $M$ contains a basis function $F_{\rm ij}(A)$, the vector $w$ contains the occupation numbers $w_{\rm ij}$ and the vector $C$ contains the CDF $C(A)$. For solving \eqref{eq:lineq}, we can use linear least squares methods in e.g. Matlab, along with regularization and a positivity constraint that $w_{\rm ij} \ge 0$. With the weights $w_{\rm ij}$ solved, we have the proportion of each bin $((b/a)_{\rm i}, \beta_{\rm j})$.

With the joint distribution for $b/a$ and $\beta$ obtained, we can compute the marginal DFs $f_{b/a}$ and $f_{\beta}$ for both parameters:
\begin{equation}
f_{(b/a)_{\rm i}} = \sum_{\rm j=1}^{l} w_{\rm ij} , \quad f_{\beta_{\rm j}} = \sum_{\rm i=1}^{k} w_{\rm ij} .
\label{eq:marginal}
\end{equation}
In addition, we can compute the CDFs for the marginal DFs. Let us denote the CDFs as $F_{b/a}$ and $F_{\beta}$. Now, assume we have obtained these CDFs for two subpopulations, $S_{1}$ and $S_{2}$ (the CDFs are denoted as $F_{b/a}(S_1)$, $F_{b/a}(S_2)$, $F_{\beta}(S_1)$ and $F_{\beta}(S_2)$), and we want to measure statistical differences of the populations. Some of such measures were used in \citet{2017A&A...601A.139N}:
\begin{equation}
 D_{b/a}(S_{1},S_{2})=\alpha_{k}\,\lVert F_{b/a}(S_{1})-F_{b/a}(S_{2})\rVert\,_{k}\,,
 \label{d_ba}
\end{equation}
\begin{equation}
 D_{\beta}(S_{1},S_{2})=\alpha_{k}\,\lVert F_{\beta}(S_{1})-F_{\beta}(S_{2})\rVert\,_{k}\,,
\label{d_beta}
 \end{equation}
where $k=1;2;\infty$ and $\alpha_{k}$ is a norm-based scaling factor: $\alpha_{1}=1/4$; $\alpha_{2}=1$ and $\alpha_{\infty}=2$. Each norm provides different kind of information about the statistical differences of the populations. The case $k=\infty$ corresponds with the Kolmogorov-Smirnov test (for details see Nortunen et al. \citeyear{2017A&A...601A.139N} or Nortunen \& Kaasalainen \citeyear{2017A&A..N}). As a general rule of thumb, two distributions can be considered significantly different statistically if $D \gtrsim 0.2$. However, a visual inspection on the marginal DF and CDF plots is also recommended for obtaining a better understanding of the statistical differences. The detailed description of the LEADER software can be found in \citet{2017A&A..N} and the software itself is available in DAMIT database\footnote{http://astro.troja.mff.cuni.cz/projects/asteroids3D/web.php}.

\section{Data}
The $1.8$-meter Pan-STARRS1 survey telescope (Hodapp et al. \citeyear{2004AN....325..636H}; Tonry et al. \citeyear{2012ApJ...750...99T}), build atop of Haleakala, Maui, started its 3-year science mission in May 2010. Photometric data were obtained in six optical and near-infrared filters ($g$, $r$, $i$, $z$, $y$ and $w$). Due to the distinct survey goals and patterns, most of the asteroids were observed in a wide-band $w$-filter ($\sim 400-700$ nm). We used the unpublished high-precision calibrated chip-stage photometry (Schlafly et al. \citeyear{2012ApJ...756..158S}) with photometric errors and selected detections of a good photometric quality. Only PSF-like and untrailed detections were considered. Our subset spanned from April 11, 2011 until May 19, 2012. In total, we had photometric data for 348 210 asteroids with about $20$ measurements for an asteroid on average. The second highest number of measurements is in the $i$-band, where we have data for 136 463 asteroids. Only the $w$-band data provided enough measurements for a reasonable application of our model. We shortly discuss results from the $i$-filter and compare them with results from the $w$-filter in Sect. \ref{i_filter}. 

\begin{figure}
\centering
\includegraphics[width=8cm]{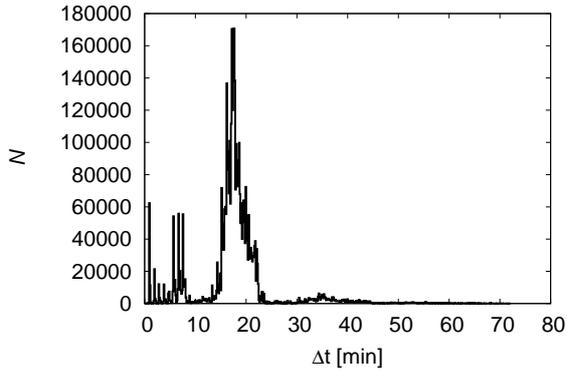}
\caption{The histogram of time intervals between measurements in the $w$-band filter from the Pan-STARRS1 survey.}
\label{hist_JD}
\end{figure}

\begin{figure}
 \centering
 \includegraphics[width=7cm]{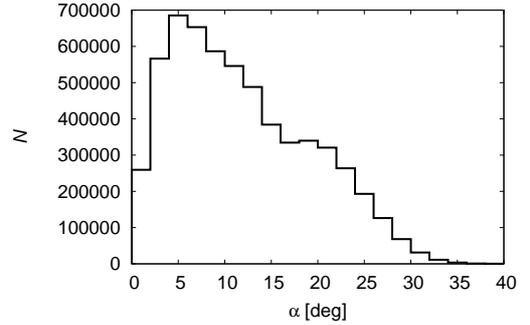}
 \caption{The histogram of the solar phase angle $\alpha$ for measurements in the $w$-filter.}
 \label{hist_phase_angle}
\end{figure}

The typical time interval between two measurements in the $w$-band filter is $\sim17$ minutes (see Fig. \ref{hist_JD}). However, not all the data were applicable to our model. Our conditions on the data were the following:
\begin{enumerate}
 \item The time interval between measurements is greater than 0.01 day ($\sim 14$ minutes). In the case of a shorter interval the rotational period would not be randomly sampled over one rotation of $\sim$hours, and in the case of a longer minimum interval we would lose a significant amount of data, as we can see from Fig. \ref{hist_JD}.
 \item Then, we limited the solar phase angle $\alpha$ to be $\leq 20^\circ$. In the model we assume this angle to be close to zero, however, in the data, there are not enough measurements with $\alpha\sim0^{\circ}$, therefore, we have to choose some reasonable value (see also Fig. \ref{hist_phase_angle}). As described in \citet{2017A&A...601A.139N} (they used $\alpha\leq30^{\circ}$) the error caused by this condition is negligible.
 \item Finally, we required at least five measurements satisfying previous conditions within 3 days to keep the geometry of observation sufficiently constant (this is the same condition as in Nortunen et al. \citeyear{2017A&A...601A.139N}).
\end{enumerate}
It is possible that for the same asteroid we had two (or even more) sets of measurements. In that case, each set was incorporated in the model.

\section{Synthetic simulations for accuracy estimation}
\label{synth_sim}
Before we compute the solution of the inverse problem from Eq.~\eqref{eq:lineq} for any Pan-STARRS1 subpopulation, we should perform a thorough analysis on whether the method is reliable and accurate with the given database. To do this, we use synthetic data created according to the procedure described in \citet{2017A&A..N}. We chose a peak of the $(b/a, \beta)$ distribution. For each asteroid in the considered population we chose a shape model from DAMIT, with $|b/a_{\rm DAMIT}-b/a_{\rm wanted}|\leq0.075$. The rotation period was chosen randomly between 3 and 12 hours from a uniform distribution (we did not use rotation periods from DAMIT, as they could be biased). Next, we used the real Pan-STARRS1 geometries and times of observations and computed the synthetic brightness using a combination of Lommel–Seeliger and Lambert scattering laws. To simulate noise, we added a minor Gaussian perturbation $1-2\%$. Our aim was to find how well the solution distribution computed from Eq.~\eqref{eq:lineq} coincides with the known, synthetic distribution. For simplicity, we are interested in reconstructing the highest peak of the joint $(b/a, \beta)$ distribution. The peak is defined as the bin with the highest occupation numbers. If there are any obvious systematic errors in the computed solution, we may attempt to apply a posterior correction to the solution. Similar synthetic simulations were used by \citet{2017A&A...601A.139N} to estimate the accuracy of the method for the WISE database, and to create a ``deconvolution'' filter to the contour image of the solution.

\subsection{Number of bodies in a population}
\begin{figure}
\centering
\includegraphics[width=0.242\textwidth]{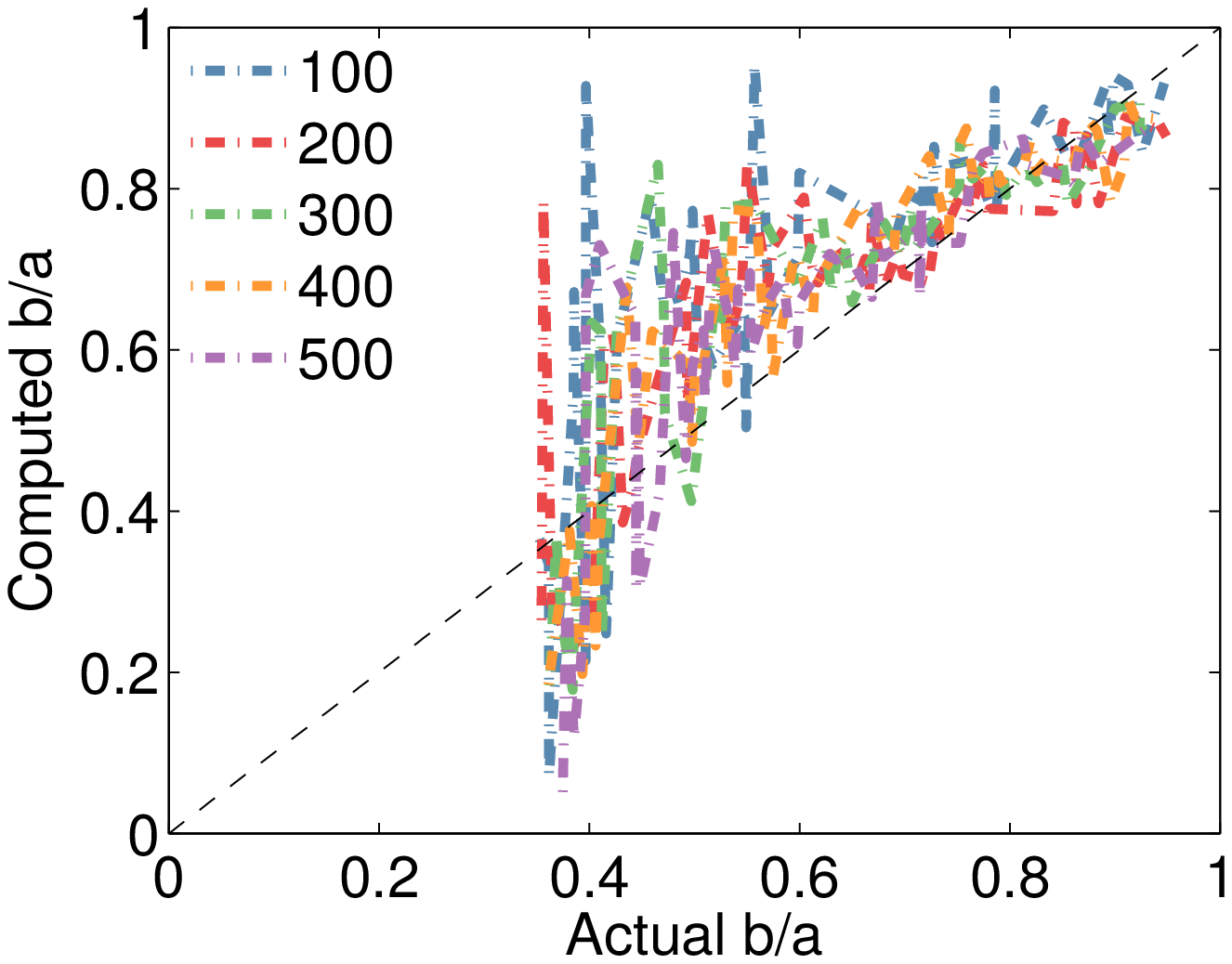}
\includegraphics[width=0.242\textwidth]{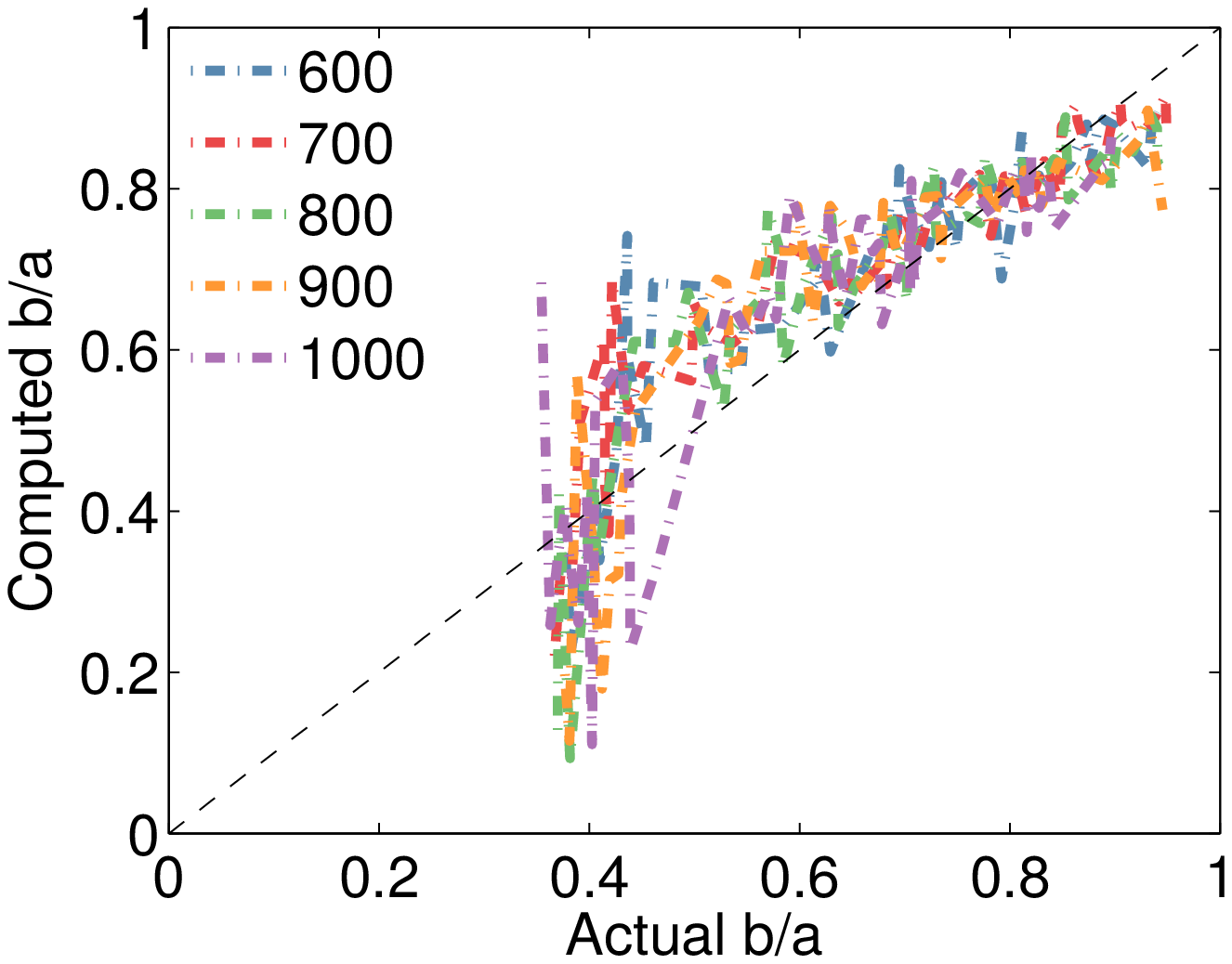}
\includegraphics[width=0.242\textwidth]{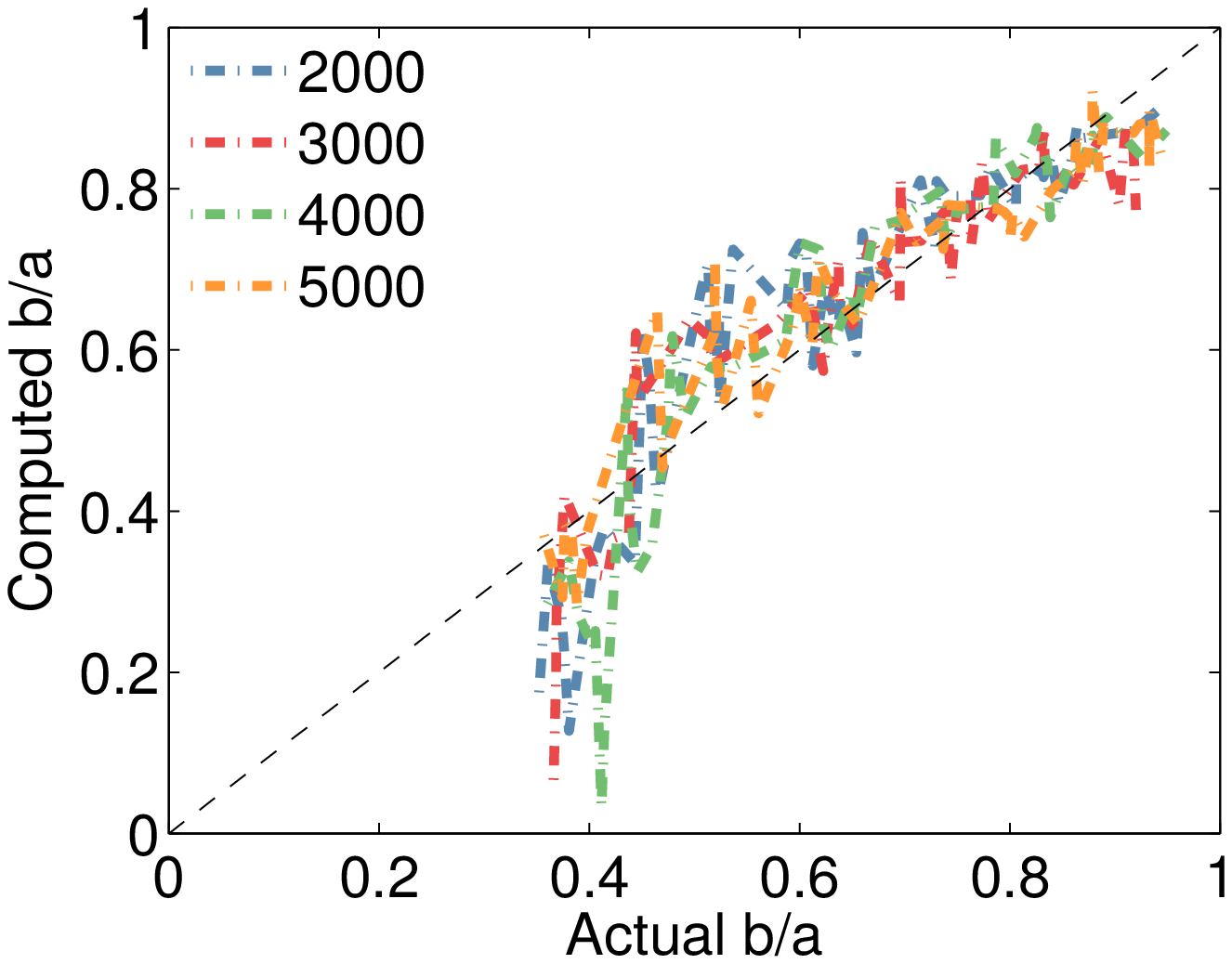}
\caption{Synthetic simulations showing how the accuracy of our method improves for $b/a$ with a growing number of asteroids (from 100 to 5000). The plots have the real peak of the distribution plotted versus the computed peak. The black dashed line of the form ``$y=x$'' depicts the ideal situation when the actual and computed peaks are the same.}
\label{synth-100-5000-p}
\end{figure}

\begin{figure}
\centering
\includegraphics[width=0.242\textwidth]{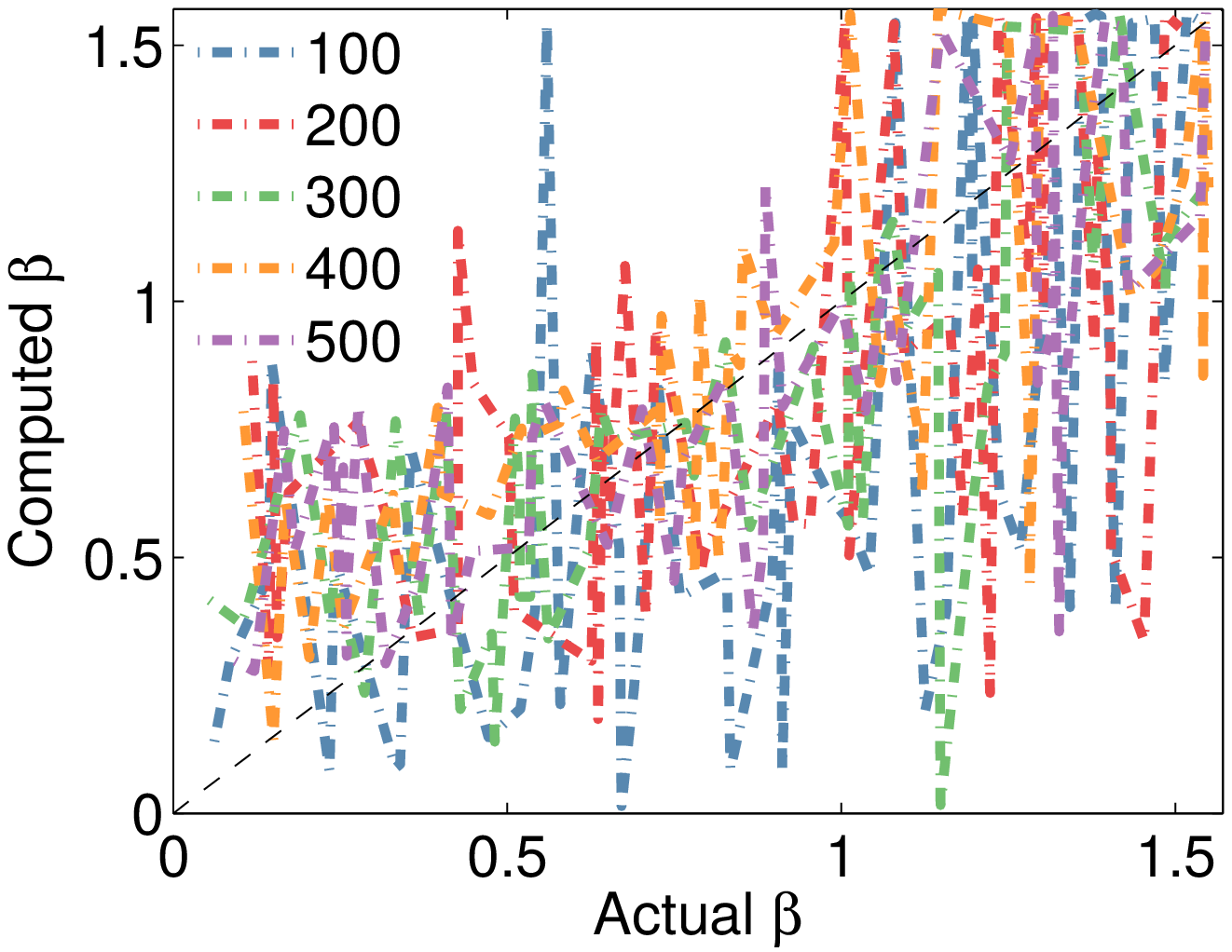}
\includegraphics[width=0.242\textwidth]{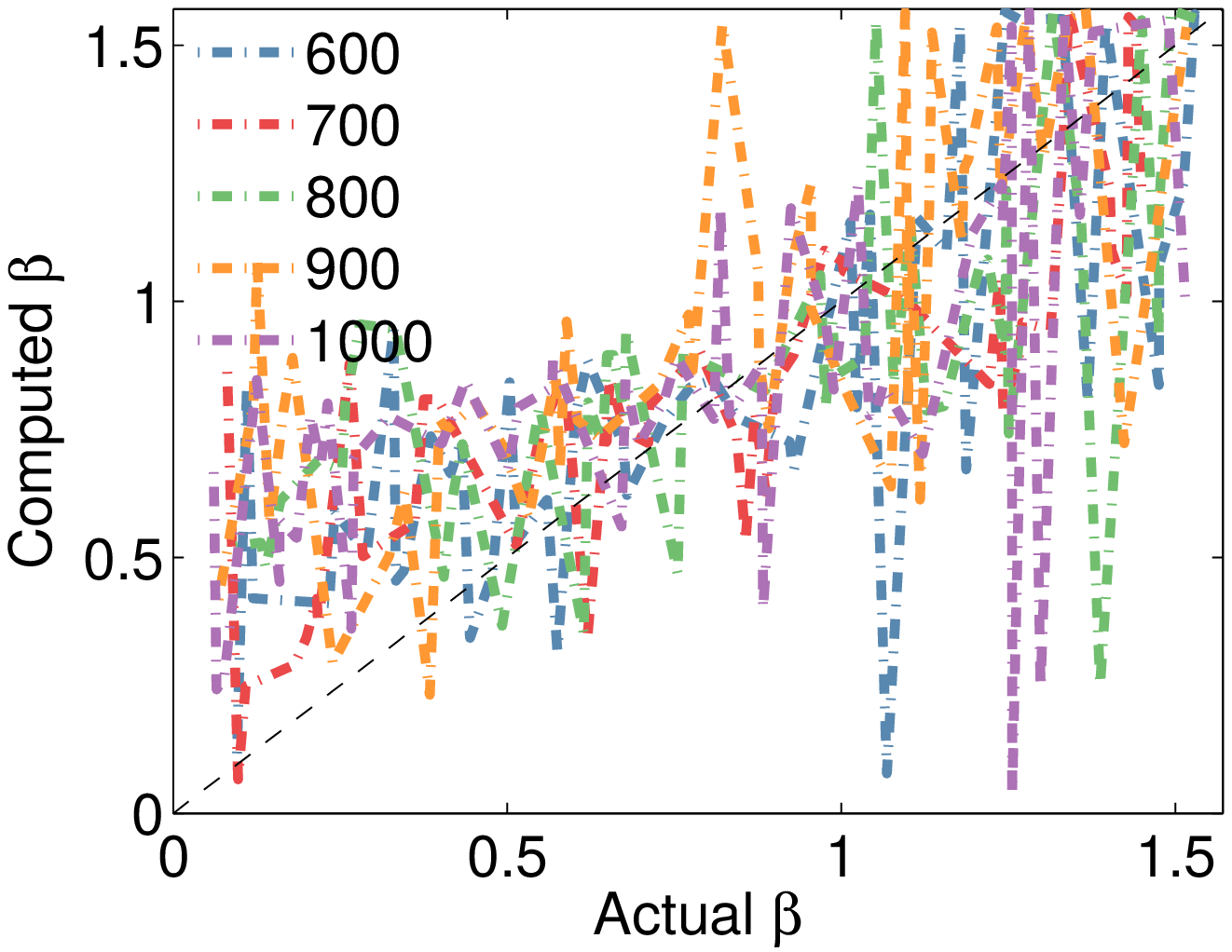}
\includegraphics[width=0.242\textwidth]{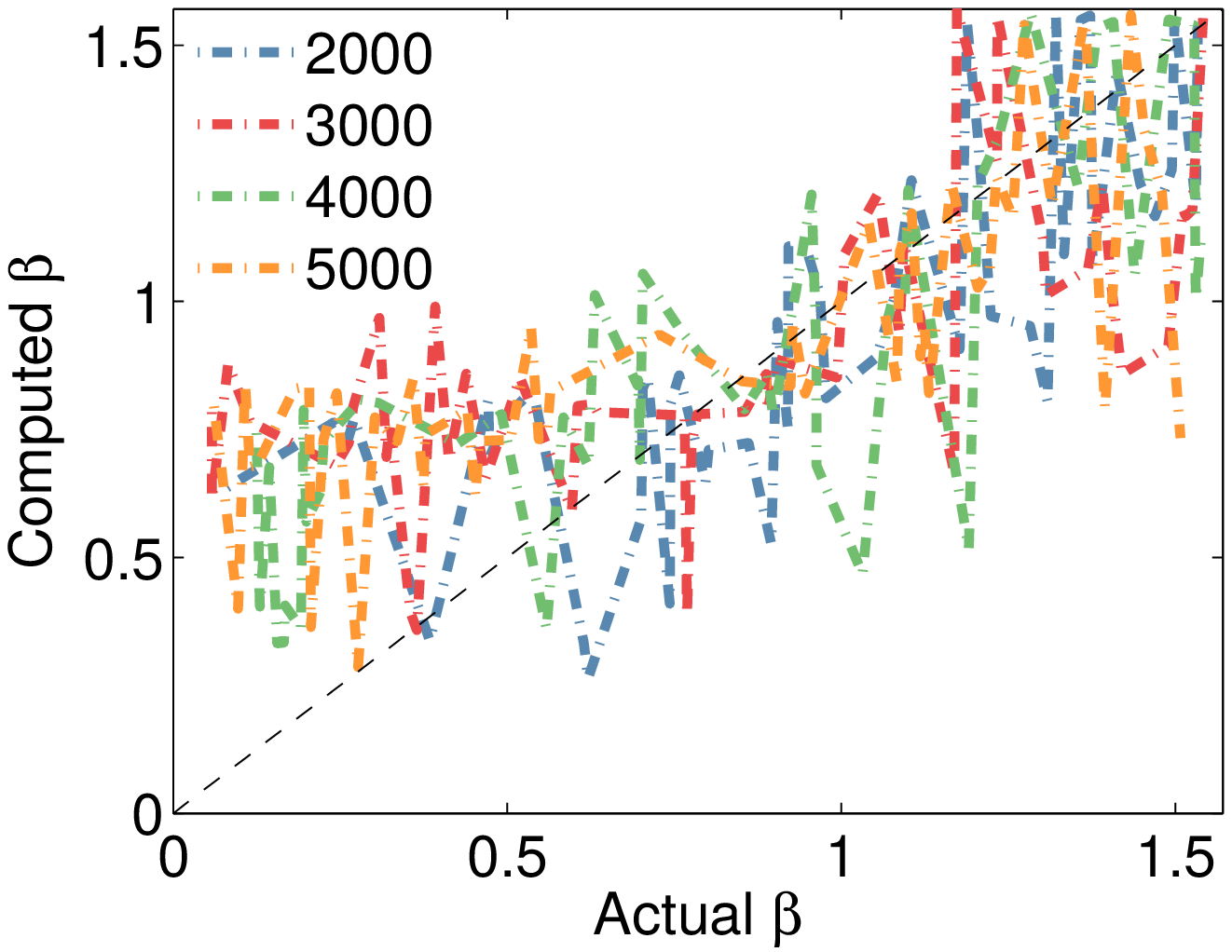}
\caption{Synthetic simulations similar to Fig.\ \ref{synth-100-5000-p}, but for $\beta$.}
\label{synth-100-5000-beta}
\end{figure}

We create 50 synthetic populations, each containing $N$ asteroids, and each population having a distinct, single peak chosen randomly. With each population, we plot the actual $(b/a, \beta)$ peak versus the computed $(b/a, \beta)$ peak to see how well they coincide. As for the populations, we set them to have from 100 to 5000 asteroids. This is so that we can evaluate how the accuracy of our method increases with a growing number of asteroids. The results from these simulations have been plotted on Fig.\ \ref{synth-100-5000-p} for $b/a$, and Fig.\ \ref{synth-100-5000-beta} for $\beta$.

As we can see from the $b/a$ plots in the left column of Fig.~\ref{synth-100-5000-p}, the accuracy of the obtained $b/a$ distribution is improved substantially as the population size increases. There is always some overshoot and undershoot when $b/a \lesssim 0.4$, but this is a rare problem: with real data, we typically have $b/a \gtrsim 0.5$, so the peak of the distribution can also be expected to be above 0.5. With a population of less than 1000 asteroids, there is a slight overshoot when $b/a > 0.5$ (i.e. the solution suggests the shapes are slightly more spherical than what they actually are), but as the population size exceeds 1000 asteroids, the computation of the $b/a$ peak is very accurate when $b/a > 0.5$.

Unfortunately, much of the $\beta$ information is lost in the inversion done for the Pan-STARRS1 database, as seen on Fig.\ \ref{synth-100-5000-beta}. For a population of less than 500 asteroids, no actual information can be recovered. For 600--1000 bodies, there is a slight increase in accuracy, but overall, the solution is too noisy to provide accurate information on $\beta$. The improvement of the accuracy is noticeable for populations with 2000--5000 asteroids, and the method provides a rough estimate on where the peak is: when the $\beta$ peak is low (perpendicular to the ecliptic plane), the obtained solution also has a low $\beta$ peak, and respectively for a high $\beta$ peak (bodies in the ecliptic plane). With the Pan-STARRS1 database, our assumption about the majority of the orbits being in the ecliptic plane may not hold well, which considerably reduces the accuracy of the beta distribution. Due to the low accuracy of the $\beta$ solution, we recommend that caution is used when interpreting the computed $\beta$ distribution. At best, our method can provide a coarse approximation on where the $\beta$ peak is located.

\begin{figure}
\centering
\includegraphics[width=0.242\textwidth]{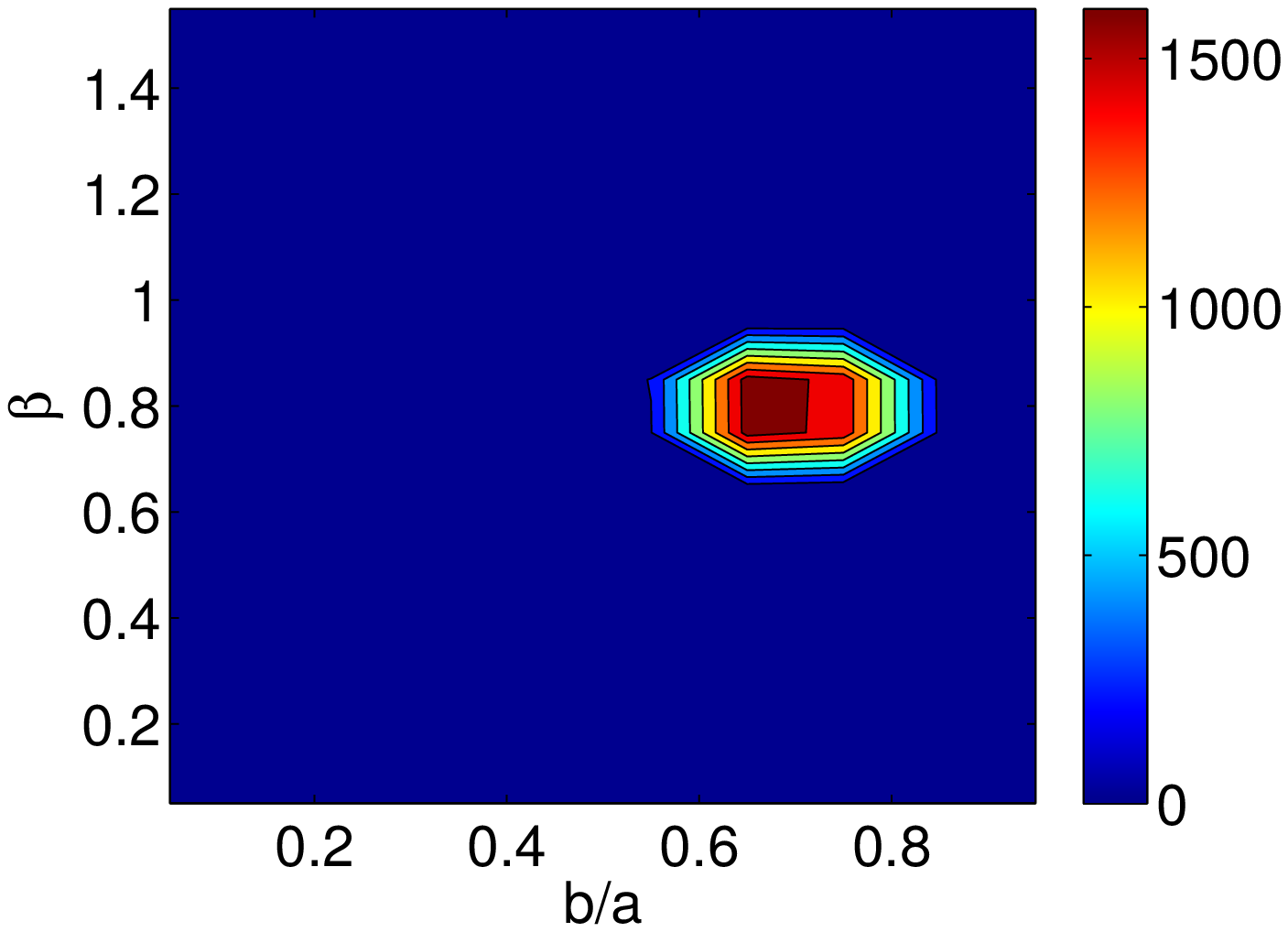}
\includegraphics[width=0.242\textwidth]{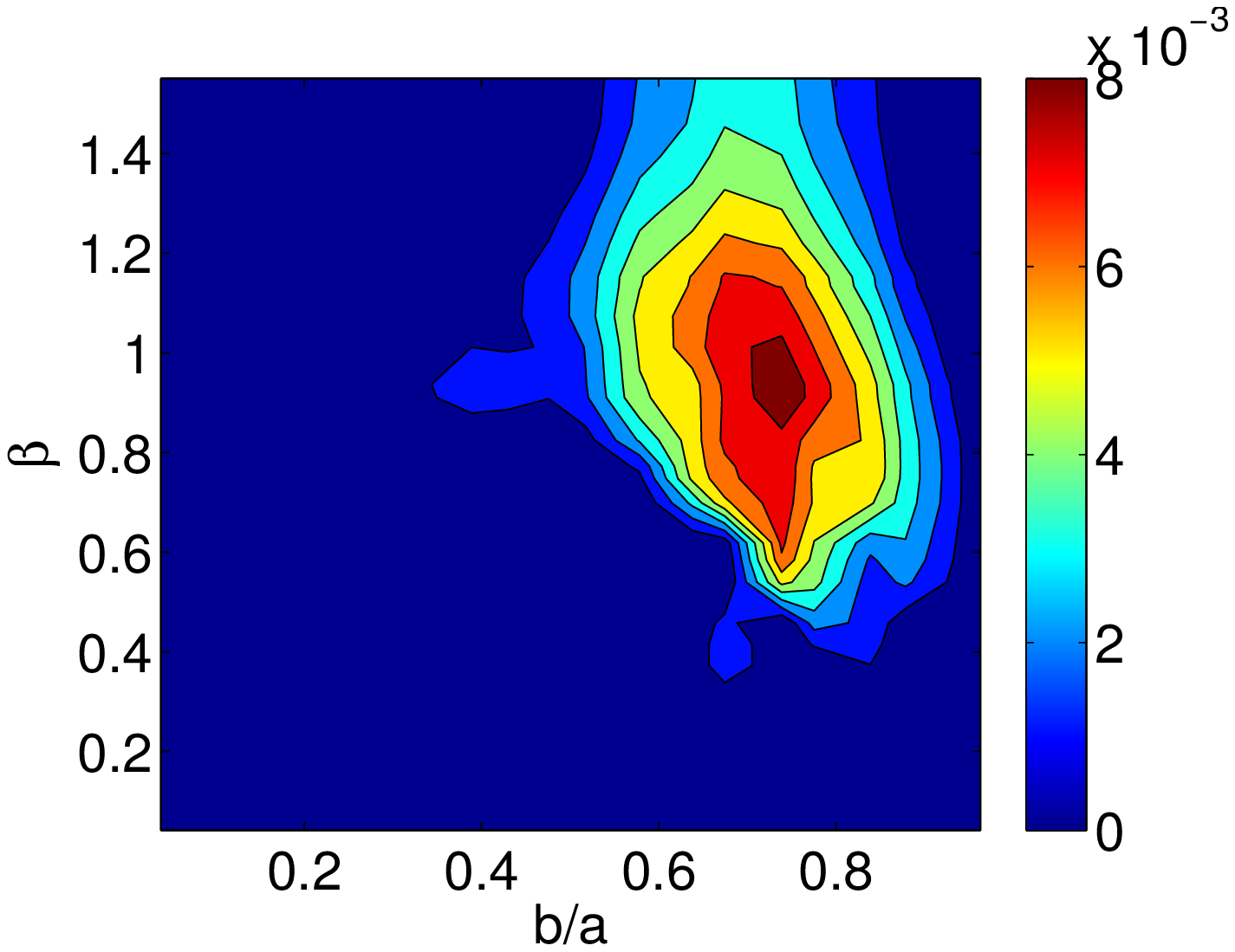}
\includegraphics[width=0.242\textwidth]{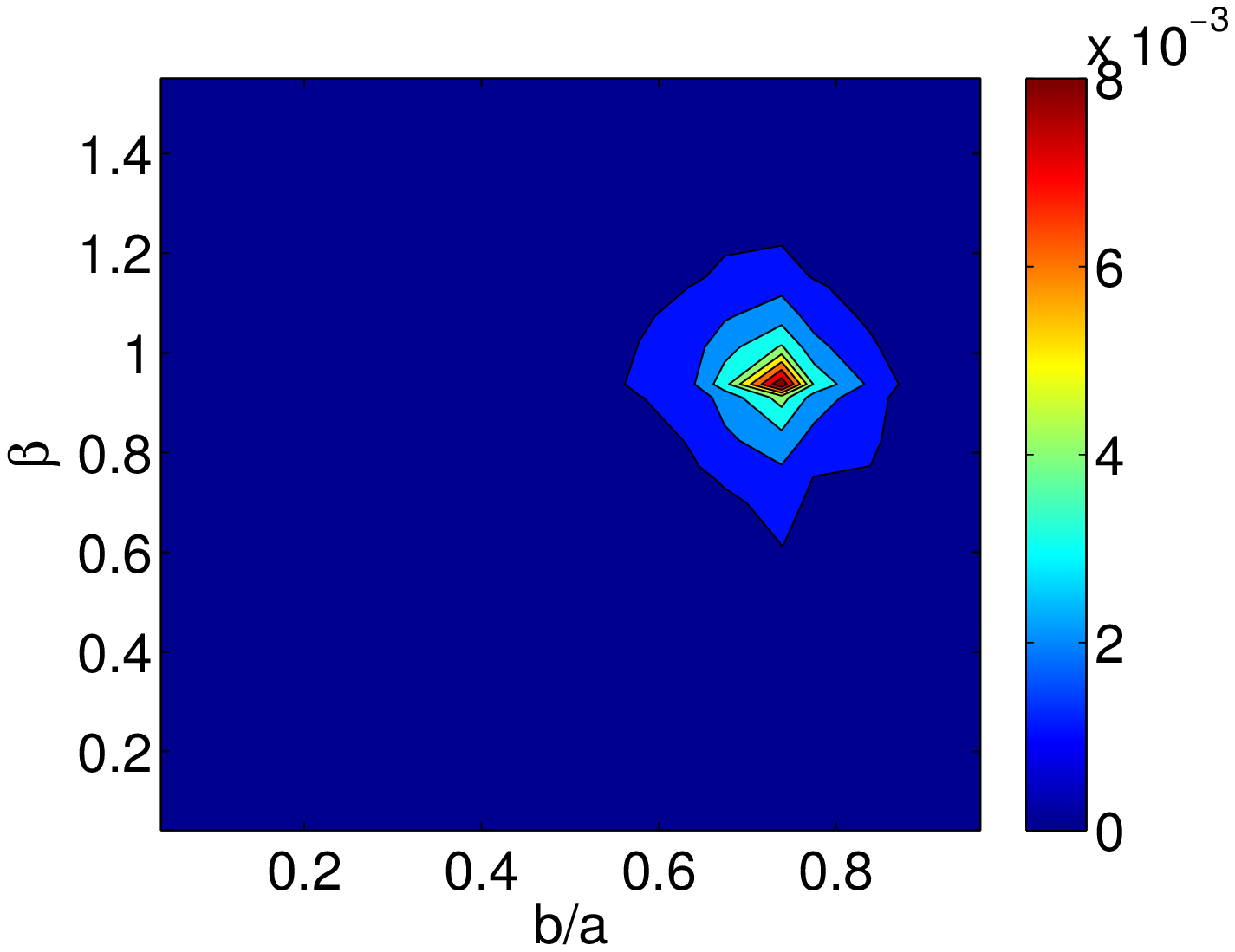}
\caption{Synthetic simulations using a fixed $(b/a, \beta)$ peak for a population of 10,000 asteroids. Top plot shows the actual $(b/a, \beta)$ distribution, middle shows the computed $(b/a, \beta)$ distribution, and bottom shows the middle solution with a deconvolution filter applied.}
\label{synth-contour}
\end{figure}

In addition to determining the correct position of the peaks, we are interested in the overall shape of the joint distribution. It is a typical tendency that the computed distribution spreads too much, especially in $\beta$ direction, and the distribution has a heavy tail towards the spin directions in the ecliptic plane. To correct this error, we may apply a deconvolution filter to the computed distribution. In this post-solution correction, we introduce some dampening by reducing the occupation numbers of bins when moving further away from the highest peak, that is, the bin with the biggest occupation number. A similar method was used in \citet{2017A&A...601A.139N}. An example of a typical solution and the effects of deconvolution has been plotted in Fig.~\ref{synth-contour}. In the simulation, we used a single, fixed $(b/a, \beta)$ peak for a population of 10 000 asteroids, with the geometries from Pan-STARRS1 database. We note that we only reduced the spreading of the solved distribution in the post-solution correction; we did not shift the position of the $(b/a, \beta)$ peak.

We emphasize that the accuracy of the solution has a strong dependence on the asteroid database used. Our method should never be used as a ``black box'' for a database. Instead, whenever we begin to utilize a new database, we should always test the validity of our method by using synthetic simulations. As the level and distribution of noise in the database is rarely known, the synthetic simulations are typically the only way to estimate the error of our method. For comparison, we performed similar synthetic simulations for WISE database in \citet{2017A&A..N}, and the results obtained from WISE and Pan-STARRS1 databases are considerably different.

\subsection{The influence of the rotation period.}
\label{influence_p}
\begin{figure}
\centering
\includegraphics[width=0.242\textwidth]{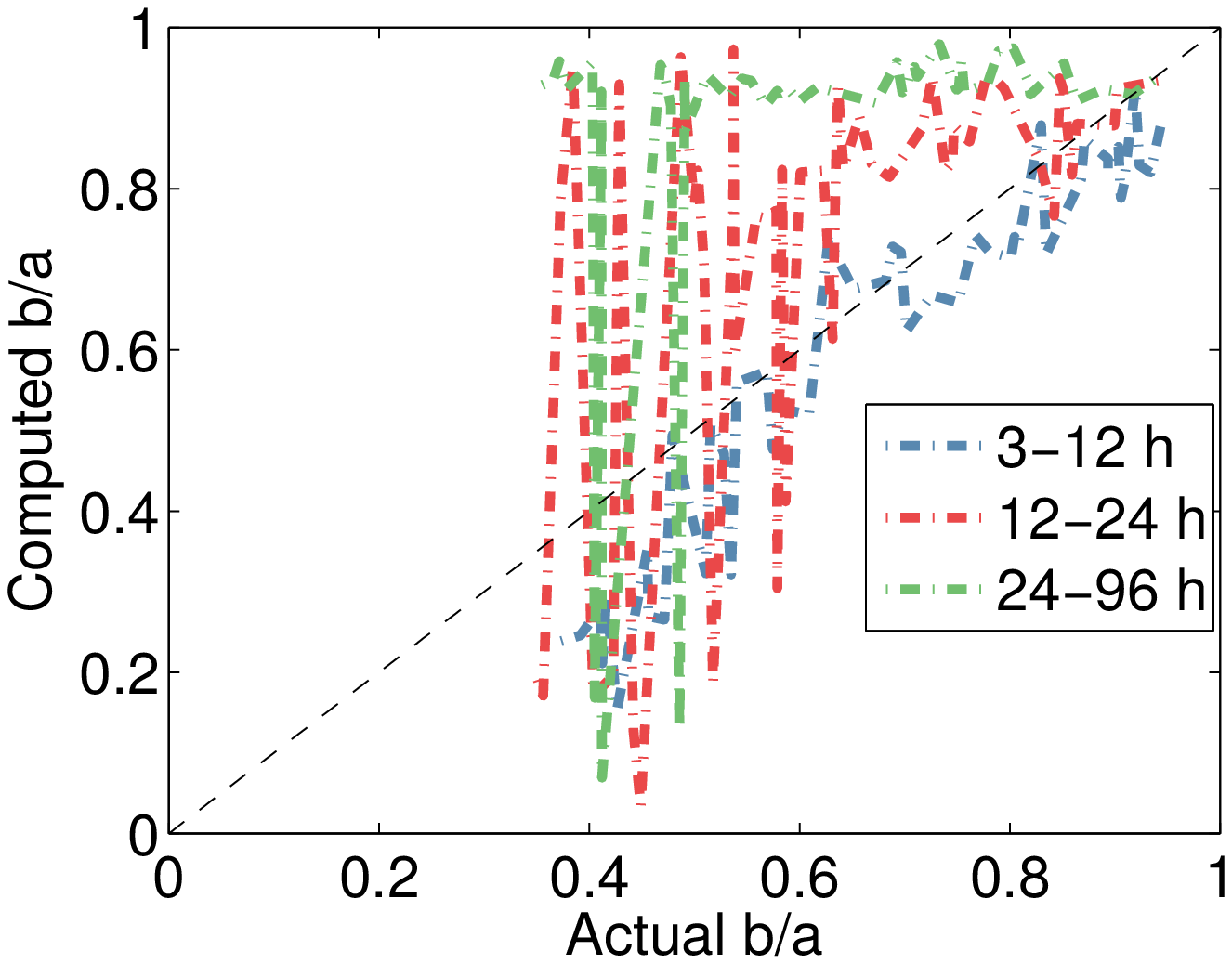}
\includegraphics[width=0.242\textwidth]{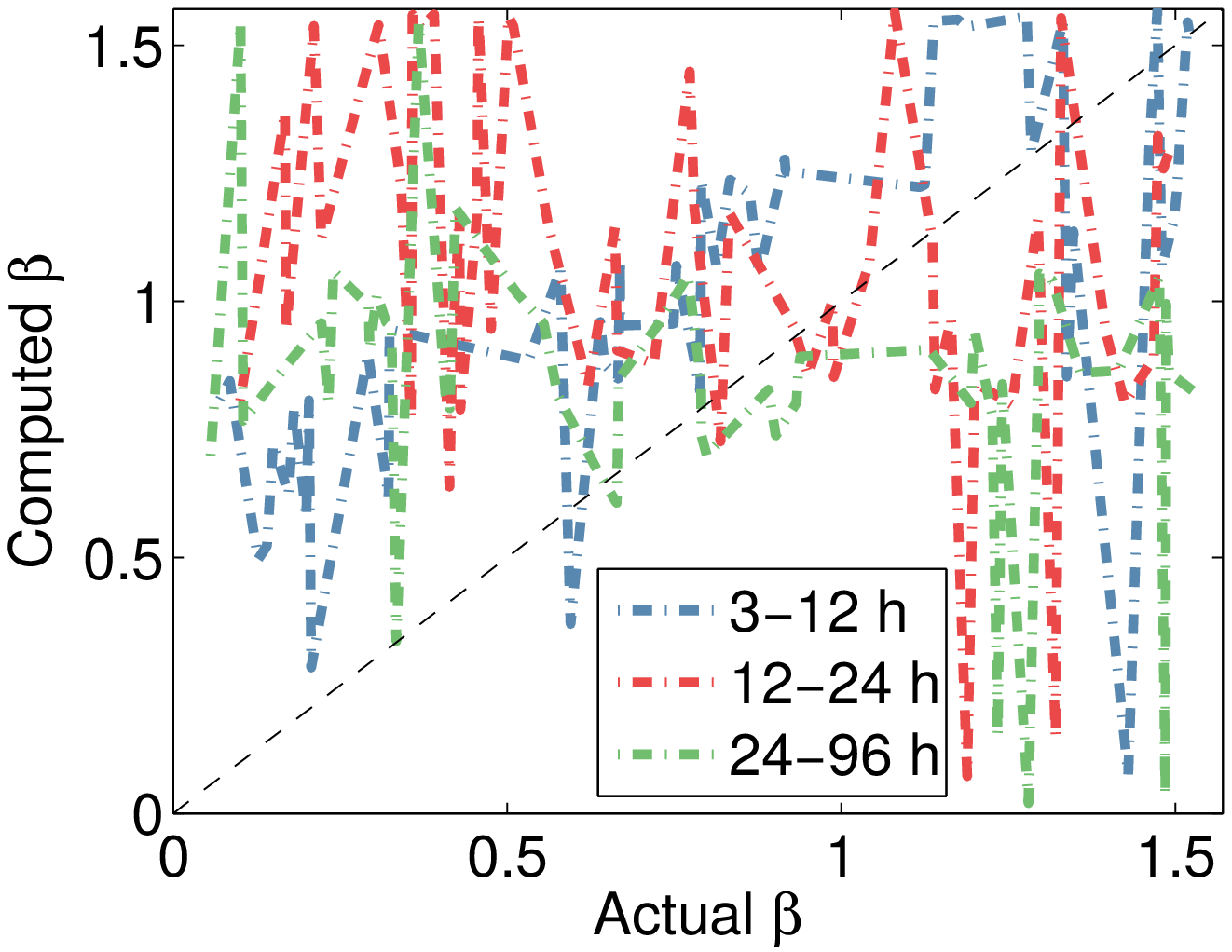}
\caption{Synthetic simulations showing the accuracy of our method for different values of rotation period $P$. The black dashed line denotes the ideal situation.}
\label{synth_rot_period}
\end{figure}

Next, we studied how accurately we are able to reproduce the known $(b/a, \beta)$ distribution when we create synthetic data by using different rotation periods $P$. We chose the following intervals of~$P$: (i) $3-12$ hours; (ii) $12-24$ hours; and (iii) $24-96$ hours. The synthetic populations contained $2\,000$ asteroids each. The results are plotted in Fig. \ref{synth_rot_period}. Considering the $b/a$ distribution, for $P<12\,$h our method provides reliable results. For $P>12\,$h the solution prefers values of $b/a\sim 1$ (spheroidal bodies) and moreover, the solution becomes unstable for $b/a<0.6$. As to the $\beta$ distribution, for $3<P<12\,$h we can notice a correlation between actual and computed $\beta$, but for $P>12\,$h, the $\beta$ is too unstable to recover any accurate information about the distribution.

The fact, that our computed distributions of $b/a$ for slow rotators ($P>12\,$h) peak at $b/a\sim 1$ is probably due to the time distribution of Pan-STARRS1 measurements. For most asteroids, data were obtained during a single night, i.e., few hours. If the real $P$ is much longer, the data cover only a small fraction of the full lightcurve (showing the time evolution of brightness during the whole $P$). The changes of brightness are thus small and our model interprets them as belonging to a spheroidal asteroid. If we construct the distribution of $P$ from the Asteroid Lightcurve Database\footnote{http://www.minorplanet.info/lightcurvedatabase.html} (LCDB, Warner et al. \citeyear{2009Icar..202..134W}) for the asteroid included in Pan-STARRS1 database we found that most of the asteroids have $P\lesssim15\,$h. Nevertheless, we have to mention that the sample of objects in the LCDB database is biased and the number of slow rotators is underestimated since it is observationally difficult to determine long periods (Marciniak et al., \citeyear{2015P&SS..118..256M}; Szab\'{o} et al., \citeyear{2016A&A...596A..40S}).

\subsection{The influence of the orbit inclination.}
Finally, we tested the influence of the orbit inclination $\sin I$ on our solution since in the model we assume $\sin I=0$. When creating the synthetic data, we used Pan-STARRS1 geometries of $2\,000$ asteroids with $\sin I\leq0.2$, i.e. first population, and $2\,000$ asteroids with $\sin I>0.2$, i.e., second population. The resulting distributions of $b/a$ and $\beta$ are not statisticaly different for populations with small and high inclinations of orbits. For $b/a$, the computed peak corresponds with the actual peak, but for $\beta$, we can notice the same problem as in Fig. \ref{synth-100-5000-beta}, the model shifts the peak to middle values.

\section{Distributions of the ratio of axes $b/a$}
In this section, we first test how many asteroids have to be in a studied subpopulation to obtain reliable results, because typically we compare subpopulations that contain different numbers of asteroids. Then we will construct the distributions of shape elongation $b/a$ for various subpopulations of main-belt asteroids. Specifically asteroids with different diameters, different rotation periods, dynamical families, taxonomic classes and subpopulations of asteroids located in different parts of the main belt (as in Cibulkov\'{a} et al. \citeyear{2016A&A...596A..57C}). To compare the distributions, we calculated $D_{b/a}$ and $D_{\beta}$ according to Eqns. (\ref{d_ba}) and (\ref{d_beta}). The bins in the distributions of $b/a$ and $\beta$ are chosen randomly, hence, for each two subpopulations that were compared, we processed ten runs and obtained 10 values of $D_{b/a}$ and $D_{\beta}$, from which we calculated the mean values. For the distribution of $b/a$ we chose 14 bins from 0 to 1, however, because the shape elongation $b/a<0.25$ is improbable, there was only one bin from 0 to 0.25, then one bin from 0.25 to 0.4 and 12 bins from 0.4 to 1. For the distribution of $\beta$ we chose 20 bins from 0 to $\pi/2$, specifically, 15 bins for $\beta > 43.4^{\circ}$ and then always one bin in following intervals: $37.2^{\circ}-43.4^{\circ}$, $31^{\circ}-37.2^{\circ}$, $24.7^{\circ}-31^{\circ}$, $18.5^{\circ}-24.7^{\circ}$ and $0^{\circ}-18.5^{\circ}$ to consider that the distribution pole latitudes is uniform in $\sin\beta$.

\subsection{The effect of the number of asteroids in a subpopulation}
\label{test_numbers}
\begin{figure}
 \centering
 \includegraphics[width=\hsize]{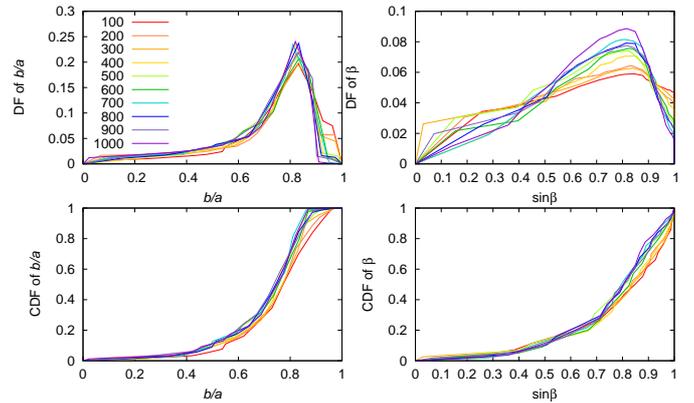}
 \caption{Distributions of $b/a$ and $\beta$ for Flora family constructed for growing number of asteroids that were included (from 100 to 1000).}
 \label{n_flora}
\end{figure}

\begin{figure}
 \centering
 \includegraphics[width=\hsize]{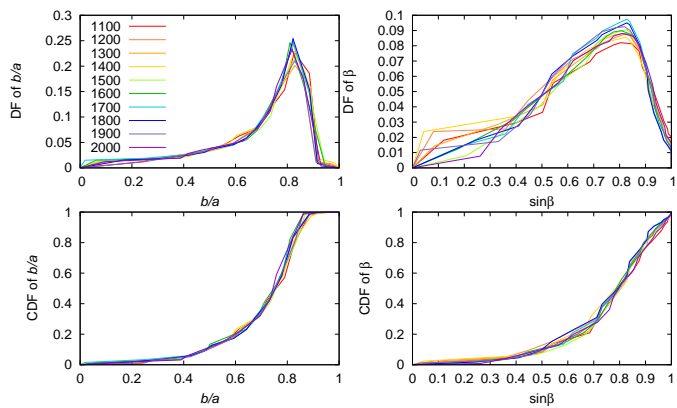}
 \caption{The same as Fig. \ref{n_flora}, but for greater number of asteroids (from 1100 to 2000).}
 \label{n_flora2}
\end{figure}

When comparing subpopulations with each other we have to take into account that they contain different numbers of asteroids. To find which population is large enough for stable results, we performed the following test: We used data for the Flora family and we randomly chose 100 of its members and ran our model ten times. We obtained ten distributions of $b/a$ and $\beta$ from which we calculated one mean distribution of $b/a$ and one for $\beta$. We repeated this for a sample of 200 randomly chosen asteroids, then 300, 400 etc., up to the sample of 2000 asteroids. All mean distributions of these subpopulations of Flora are shown in Figures \ref{n_flora} and \ref{n_flora2}. For the distribution of $b/a$ we can see that the results are stable from $\sim800$ asteroids in the subpopulation. However, for $\beta$ the results are much more unstable, the distributions are clearly different even in Fig. \ref{n_flora2} that contains populations with 1100 to 2000 asteroids. With growing number of bodies the peak of $\beta$ distribution is higher and the number of asteroids with $\beta\sim\pi/2$ decreases.

\subsection{Asteroids with different diameters}
\begin{figure}
\centering
\includegraphics[width=\hsize]{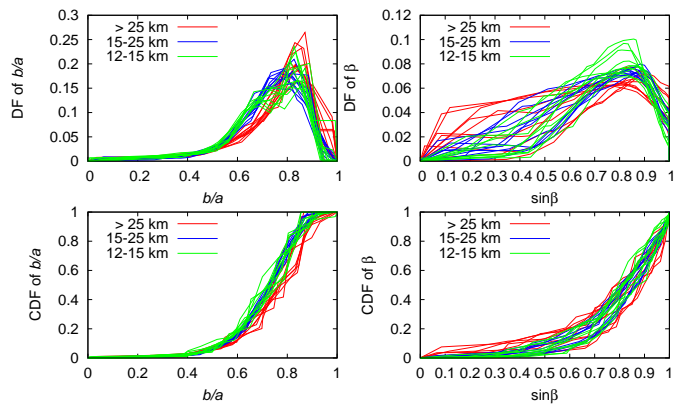}
\caption{DFs and CDFs of $b/a$ and $\beta$ for asteroids with $12<D<15$ km (green lines), $15<D<25$ (blue lines) and $D>25$ km (red lines).}
\label{sizes1}
\end{figure}

First, we focused on groups of asteroids with different diameters $D$. For asteroids which have $D$ derived from the observations of WISE satellite, we used that value, for other asteroids we used diameters from AstOrb catalog. We divided asteroids into seven groups: with $D<3\,$km; $3-6\,$km; $6-9\,$km; $9-12\,$km; $12-15\,$km; $15-25\,$km; and $D>25\,$km; and compared them with each other. For $D<15\,$km we have in all five groups more than 1200 asteroids, however, there are only 990 asteroids with $15<D<25\,$km and only 223 bodies in the last group ($D>25\,$km), which is not enough for a reliable result. The distributions for three groups with the largest $D$ are shown in Fig. \ref{sizes1}. Although, it is in agreement with the findings of \citet{2016A&A...596A..57C} that the asteroids larger than $D>25\,$km are more often spheroidal, in our case it might be just an effect of the low number of asteroids in the last subpopulation.

The mean values of $D_{b/a}$ for distributions of $b/a$ for the three subpopulations with largest diameters are listed in Table~\ref{ks_ps}. The distributions of $b/a$ for groups of asteroids with $D<15\,$km are not statistically different from the group of asteroids with $15<D<25\,$km and have a maximum for $b/a\sim 0.8$. The average axial ratio $b/a$ from Pan-STARRS1 survey was also determined by \citet{2016MNRAS.459.2964M}. For asteroids with $D<8$ km they found the average $b/a$ to be 0.85, which is a little more spheroidal than our result. For $D<25\,$km, \citet{2016A&A...596A..57C} found the maximum of distribution of $b/a$ for $\sim0.63$, i.e., asteroids are more elongated, nevertheless, the possibility is there mentioned that the results could be influenced by the underestimated data noise, which causes shape estimates to be more elongated.

\begin{table}
 \small
\centering
\begin{tabular}{cccc}
\hline
\hline
\rule{0cm}{2.5ex}
 populations & $D_{b/a}(L^1)$ & $D_{b/a}(L^2)$ & $D_{b/a}(L^{\infty})$ \\
\hline
\rule{0cm}{2.5ex}%
 $D= 15-25$ km; $>25$ km & 0.164 & 0.269 & 0.351\\
 $D= 12-15$ km; $15-25$ km & 0.091 & 0.146 & 0.189\\
 $P= 0-4$ h; $4-8$ h & 0.369 & 0.537 & 0.573\\
 $P= 0-4$ h; $8-15$ h & 0.107 & 0.163 & 0.204\\
 $P= 4-8$ h; $8-15$ h & 0.450 & 0.638 & 0.642\\
 Flora; background & 0.087 & 0.159 & 0.244\\
 Massalia; background & 0.294 & 0.462 & 0.554\\
 Nysa Polana; background & 0.140 & 0.259 & 0.399\\
 Vesta; background & 0.068 & 0.114 & 0.170\\
 Phocaea; background & 0.175 & 0.274 & 0.367\\
 Eunomia; background & 0.079 & 0.123 & 0.176\\
 Gefion; background & 0.132 & 0.203 & 0.270\\
 Maria; background & 0.078 & 0.129 & 0.186\\
 Koronis; background & 0.142 & 0.244 & 0.367\\
 Eos; background & 0.084 & 0.142 & 0.208\\
 Hygiea; background & 0.098 & 0.163 & 0.218 \\
 Themis; background & 0.134 & 0.243 & 0.342\\
 Alauda; background & 0.144 & 0.219 & 0.250\\
 C class; S class & 0.081 & 0.129 & 0.174\\
 Massalia; background ($i$-filter) & 0.273 & 0.415 & 0.495\\
 Phocaea; background ($i$-filter) & 0.212 & 0.291 & 0.333\\
 $w$-filter; $i$-filter for Nysa Polana & 0.095 & 0.160 & 0.220\\
 \hline
\end{tabular}
\vspace{2mm}
\caption{\footnotesize{The parameter $D_{b/a}$ for selected pairs of populations that were compared. The given values are the mean values from ten runs of our model.}}
\label{ks_ps}
\end{table}

\begin{figure}
 \centering
 \includegraphics[width=\hsize]{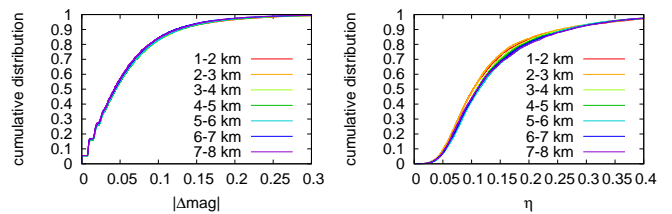}
 \caption{Left: Cumulative distributions of absolute value of change in magnitude $|\Delta\rm{mag}|$ for groups of asteroids with different sizes. Right: Cumulative distributions of $\eta$ for groups of asteroids with different sizes.}
 \label{eta_cumul}
\end{figure}

We also tried to reconstruct the cumulative distributions of absolute rate of change in magnitude from work \citet{2016MNRAS.459.2964M}, who constructed distributions for asteroids with $1<D<8\,$km and divided them into groups $1-2\,$km, $2-3\,$km etc. to $7-8\,$km. They found that with decreasing diameter, the distributions show smaller change in magnitude. However, we could not find any differences between individual distributions (see Fig. \ref{eta_cumul} on the left). The possible explanation of this disagreement is that \citet{2016MNRAS.459.2964M} used only measurements with magnitude uncertainty $\leq 0.02$, however, we used all measurements, our only conditions were (i) the solar phase angle $\alpha<10^{\circ}$ (this is the same condition as in McNeill et al., \citeyear{2016MNRAS.459.2964M}) and (ii) pairs of measurements separated by time interval $10$ min $<\Delta t <20$ min. We constructed also cumulative distributions of brightness variation $\eta$ to see if there will be any differences, but as shown in Fig. \ref{eta_cumul} on the right, the $\eta$ distributions for groups of asteroids with different diameters are almost the same.

Then we focused on the distributions of $\beta$. As we can see in Fig. \ref{sizes1} on the right, they look different from results of e.g. \citet{2016A&A...596A..57C} or \citet{2011A&A...530A.134H}, where $\beta$ is clustered around 0 due to the Yarkovsky-O'Keefe-Radzievskii-Paddack (YORP) effect, that shifts $\beta$ near the pole of the ecliptic (e.g., Pravec \& Harris \citeyear{2000Icar..148...12P}, Rubincam \citeyear{2000Icar..148....2R}). Nevertheless, as explained in Sec. \ref{synth_sim} or \ref{test_numbers}, we found that the distribution of $\beta$ is considerably influenced by the number of asteroids in given subpopulation and becomes flatter with decreasing number of asteroids. In Fig. \ref{synth-100-5000-beta} we can also see that the model tends to shift the peak to the middle values. The results on $\beta$ are thus nor reliable and in the following tests we will only focus on the distributions of $b/a$.

Because the number of asteroids with $D>25$ km in data from Pan-STARRS1 is insignificant in comparison to the number of smaller asteroids (less than $1$\%), this dependence on diameter does not influence the results of the following tests.

\subsection{Different rotation periods}
According to their rotation periods $P$ provided by the LCDB database, we divided asteroids into three groups. To ensure that all groups are populous enough for stable results we chose the following intervals: (i) $P=0-4\,$h (1081 bodies); (ii) $4-8\,$h (1967 bodies); and (iii) $8-15\,$h (1071 bodies). We excluded asteroids with $P>15\,$h since our simulations with synthetic data showed the results are not reliable (see also Fig.~\ref{synth_rot_period}).

\begin{figure}
\centering
\includegraphics[width=\hsize]{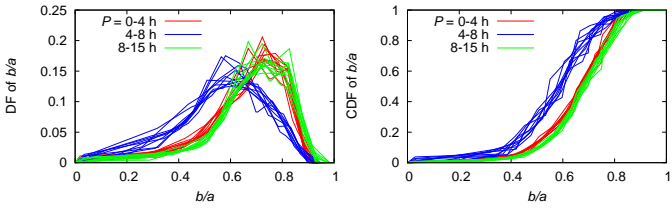}
\caption{DFs and CDFs of $b/a$ for asteroid populations with different rotation periods.}
\label{rot_period_real}
\end{figure}

We compared populations with each other and plotted their distributions of $b/a$ in Fig. \ref{rot_period_real}. We can see that the fastest rotators ($P=0-4\,$h) are on average more spheroidal than the population with $P=4-8\,$h, but their $b/a$ distribution is not different from the third population with $P=8-15\,$h. The mean values of $D_{b/a}$ are listed in Table \ref{ks_ps}.

\begin{figure}
\centering
\includegraphics[width=4.4cm]{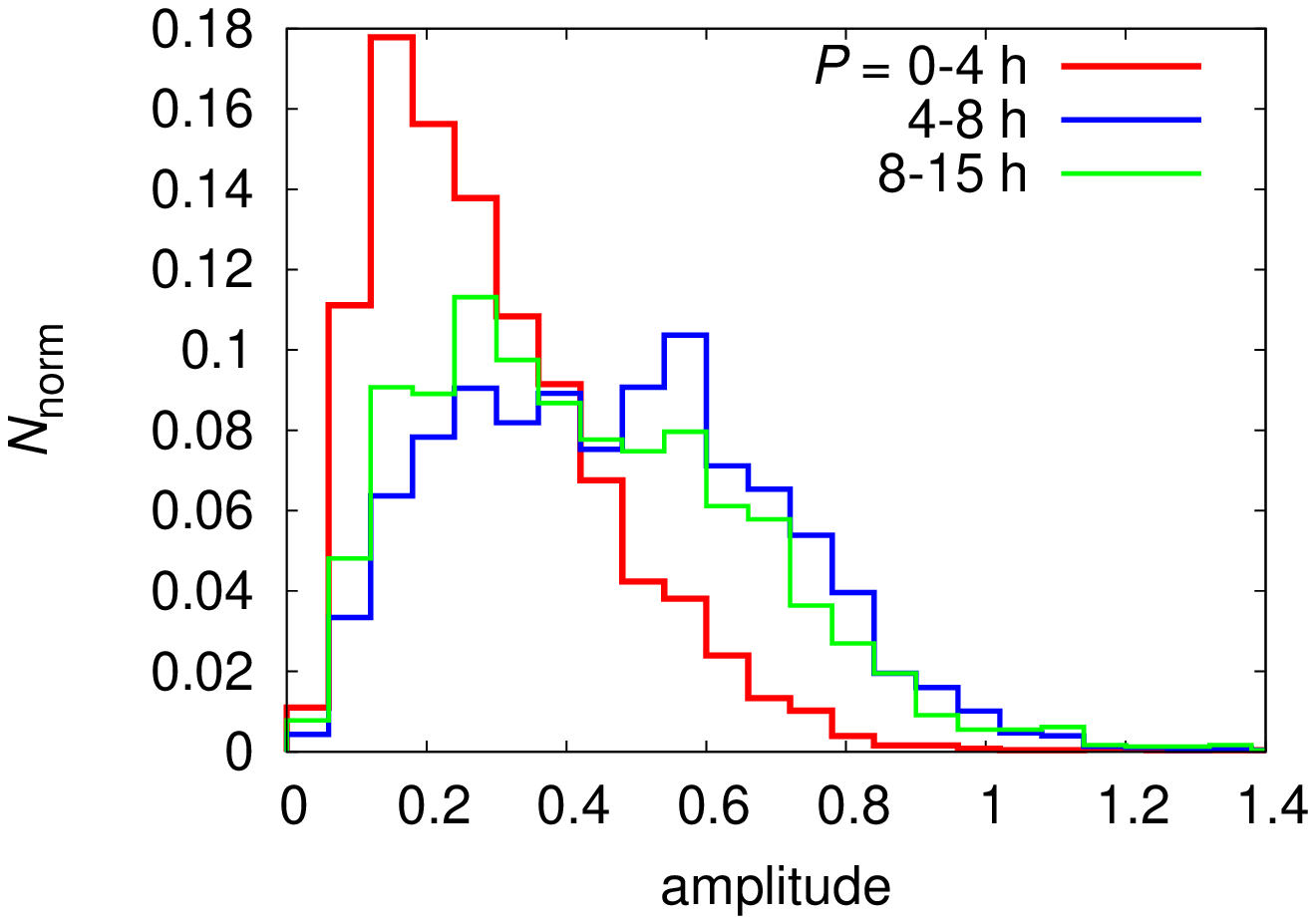}
\includegraphics[width=4.3cm]{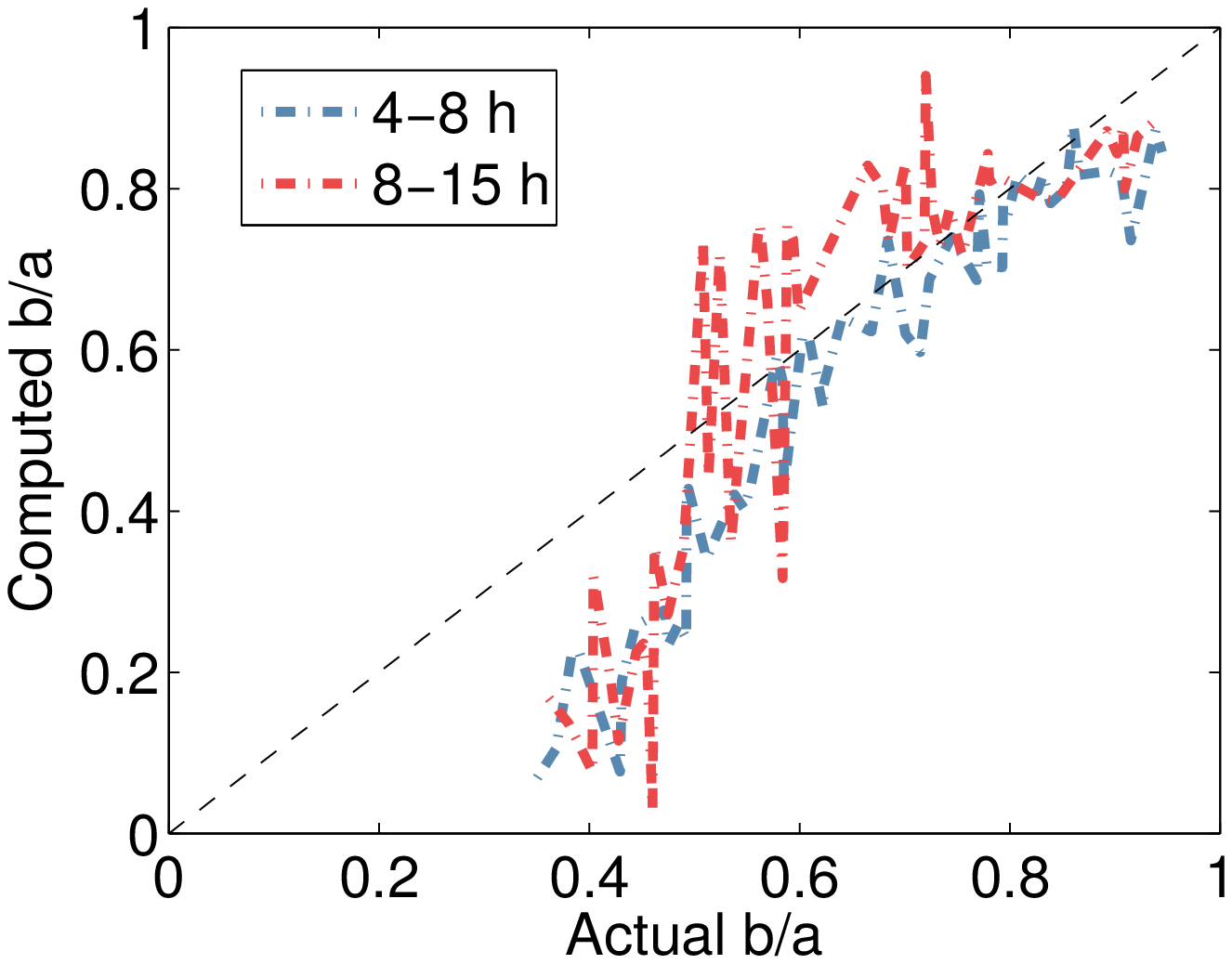}
\caption{Left: Distributions of lightcurves amplitudes from the LCDB database for different rotation periods $P$. Right: Synthetic simulations showing the accuracy of our method for two different intervals of rotation period $P$.}
\label{period_ampl}
\end{figure}

The critical rotation rate is, for the same density, dependent on the elongation (Pravec \& Harris \citeyear{2000Icar..148...12P}). The spheroidal bodies are thus able to rotate faster that the elongated ones, which is in accordance with our results for the first two populations. However, we were not able to explain why the third population, with $P=8-15\,$h, should contain more spheroidal asteroids than the population with $P=4-8\,$h. Therefore, using the LCDB database we constructed distributions of lightcurves amplitudes for the three above mentioned populations (see Fig. \ref{period_ampl}). Higher amplitudes correspond with larger elongations. The distributions of the first two groups are in accordance with the results from Pan-STARRS1 data, but for the third population ($P=8-15\,$h), we obtained similar distribution as for the population with $P=4-8\,$h.

To explain this discrepancy we performed another test with synthetic data. We used the same setup as in Sec. \ref{influence_p}, where we studied the influence of the rotation period on the accuracy of the solution, but we chose populations with $P=4-8\,$h and $P=8-15\,$h. The resulting distributions of $b/a$ are shown in Fig. \ref{period_ampl}, on the right. We can see that for both populations, our model is not able to correctly reproduce peak $b/a\lesssim 0.6$, nevertheless such elongation peak is uncommon, most of the asteroids have $b/a>0.6$. Considering $0.8>b/a>0.6$, for the population with $P=8-15\,$h, our model provides slightly more spheroidal objects ($b/a\sim b/a_{\rm actual}+0.1$) and for the population with $P=4-8\,$h, it provides slightly more elongated objects ($b/a\sim b/a_{\rm actual}-0.05$). We can conclude that the difference between $b/a$ distributions for these two populations (shown in Fig. \ref{rot_period_real}) is due to the method bias that shifts their $b/a$ values $\sim 0.15$ apart.

\subsection{Period from estimated photometric slopes}
By this analysis we have learned that our distributions of $b/a$ for other asteroid populations can be strongly influenced by the appropriate period distributions. Unfortunately, our model does not provide the rotation period $P$ and the LCDB database contains $P$ for only $\sim 14\,000$ asteroids. That sample, divided into individual populations, is not large enough for a statistical purpose. Nevertheless, we noticed that $P$ could be formally calculated directly from photometric data if there are many measurements for an asteroid and if they are appropriately distributed in time. More precisely, we need pairs of measurements close in time and also a sufficient number of such pairs. 

First we derive a general result for the time series of any signal $I$ that is of pure sinusoidal form of $n$th order only, augmented by the mean term $I_0$ (0th order): 
\begin{equation} 
I=I_0+\cos n\omega t 
\label{sinusoid} 
\end{equation} 
(we can choose this form since the starting point is irrelevant), where $\omega$ is the rotation frequency. If the estimates of the time derivative ${\rm d}I/{\rm d}t$ are available (i.e., measurements of $I$ within a short time interval as with Pan-STARRS1), we can use these to estimate $\omega$ and hence the period $P=2\pi/\omega$ in a simple manner. Using the variation (standard deviation) $\Delta$ as defined with Eq. (1), with $I=L^2$, and computing the mean $\langle\vert {\rm d}I/{\rm d}t \vert\rangle$ from Eq. (\ref{sinusoid}) by integrating ${\rm d}I/{\rm d}t$ over the interval $[0,\pi/2]$, we directly obtain 
\begin{equation} 
P=4 n \sqrt{2} \Delta(I)/\langle\vert {\rm d}I/{\rm d}t \vert\rangle\,. 
\label{period_eq} 
\end{equation} 
Since $I=L^2$ for an ellipsoid is of the pure $n=2$ double-sinusoidal form (Nortunen et al. \citeyear{2017A&A...601A.139N}), we can use Pan-STARRS1 slope estimates $\vert {\rm d}L^2/{\rm d}t \vert$ and their mean $\langle\vert {\rm d}L^2/{\rm d}t \vert\rangle$ to obtain the period with the aid of Eq. (\ref{period_eq}). However, for each asteroid this requires a number of slope estimates. The derivation ${\rm d}L^2/{\rm d}t$ can be approximately calculated from pairs of measurements close in time, but there is a lower limit due to the accuracy of data. We chose ${\rm d}t>10$ min to distinguish the change of brightness from data noise.

To verify if this relation can be used in practice, we performed a test on synthetic data created as follows: using the DAMIT models, the Hapke scattering model (Hapke \citeyear{1981JGR....86.4571H}; Hapke \citeyear{1993tres.book.....H}) with randomly chosen parameters and randomly chosen rotational period $P$ (uniformly distributed from 2 to $50\,$h), we calculated synthetic brightness that we assigned to $\sim 1000$ asteroids observed with Pan-STARRS1 (we left the geometry of observations unchanged), for which we had the largest number of measurements. From these new synthetic brightnesses we can calculate the period $P$ according to Eq. (\ref{period_eq}) that should approximate the synthetic $P$.

\begin{figure}
\centering
\includegraphics[width=\hsize]{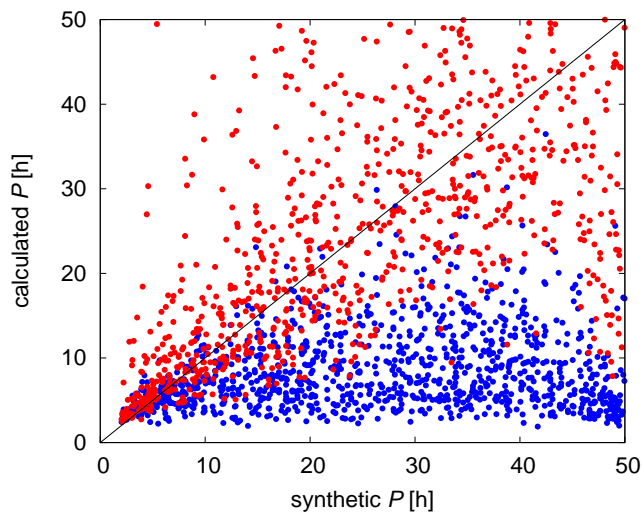}
\caption{The comparison of calculated and synthetic rotational period~$P$. The red points denote synthetic data without noise, the blue points synthetic data with noise 0.02.}
\label{period_synth}
\end{figure}

The derivative $\langle \lvert {\rm d}L^2/{\rm d}t\rvert\rangle$ was computed from pairs of measurement separated by time interval $10<\Delta t< 20$ min and we required at least 12 pairs (to calculate the mean value) within 5 days. The variation $\Delta L^2$ was also computed within 5 days. We tested synthetic data without any noise and also data with Gaussian noise of $2\%$. We compared the calculated $P$ with the synthetic one by computing the correlation coefficient: data with noise show no correlation (the coefficient is 0.19) and as we can see in Fig. \ref{period_synth} (blue points), there is a strong preference for low values of $P$. Interestingly enough, the bias is systematic and amounts to an underestimation factor of about 0.5 for the point fan. Apparently noise systematically increases the slope average from the pairwise slope estimates. The situation for data without noise is slightly better (coefficient 0.30) and if we consider only periods from interval 2 to 30 hours, the correlation coefficient is 0.65 (see also Fig. \ref{period_synth}). For periods under ten hours, the points are even more tightly clustered near the $x=y$ correlation line.

The possible reason for this bad correlation could be the insufficient number of measurements from which the mean values are calculated. Therefore, to each measurement we added two another, one 0.01 d ($14.4$ minutes) earlier and the second 0.01 d later. In total, we had three times more measurements for each asteroid. However, the resulting $P$ were not significantly different from the previous test, in the interval of $P$ from 2 to 30 hours, the correlation coefficient is $0.60$.

We also tested the relation (\ref{period_eq}) on real data from Pan-STARRS1 survey, however, there were only few asteroids for which we had required number of measurements (as described above) and at the same time also the information about the real rotational period from the LCDB Database. For these bodies we did not obtain a good agreement between the estimated and the real periods. Apparently the use of the period estimate Eq.\ (\ref{period_eq}) requires a large number of well-distributed slope pairs over a rotation cycle. Also, a low number of pairs exacerbates the effects of noise and deviations from the pure double-sinusoidal form. Estimates based on the derivative of a function are usually considerably more unstable than those based on the function itself. This approach is thus not applicable in practice and we are not able to correct $b/a$ distributions of other asteroid populations to have the same $P$ distributions.

\subsection{Dynamical families}
\begin{figure}
\centering
\includegraphics[width=9cm]{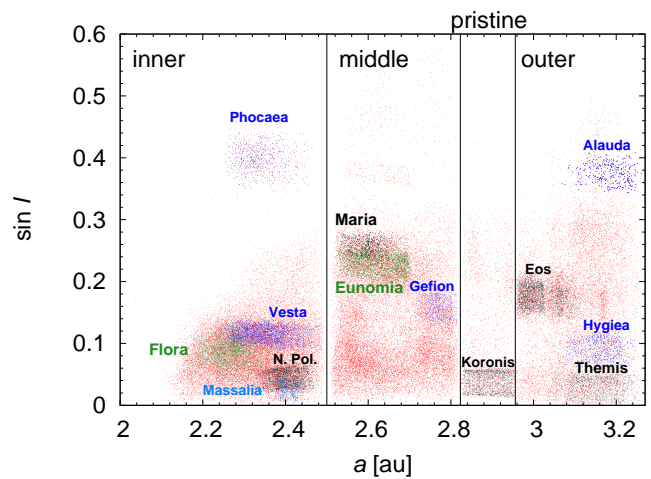}
\caption{Four parts of the main asteroid belt defined according to the proper semimajor axis $a$ (we used proper values of $a$ and $I$ from Asteroids Dynamic Site; \citep{2003A&A...403.1165K}).}
\label{5parts}
\end{figure}

\begin{figure}
\centering
\includegraphics[width=\hsize]{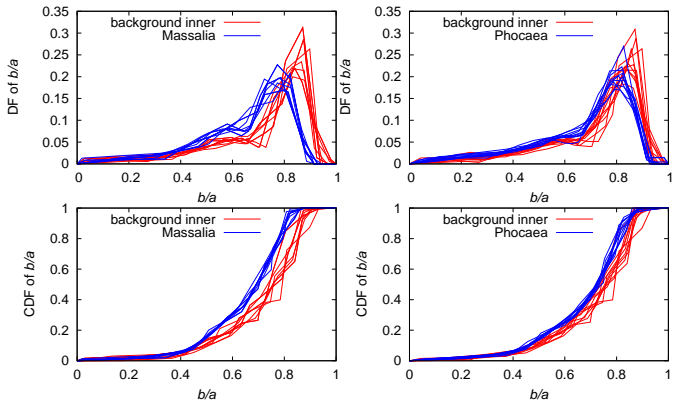}
\caption{Left: DFs and CDFs of $b/a$ for Massalia family (blue lines) and its background (red lines). Right: The same for Phocaea family.}
\label{massalia_phocaea}
\end{figure}

\begin{figure}
\centering
\includegraphics[width=\hsize]{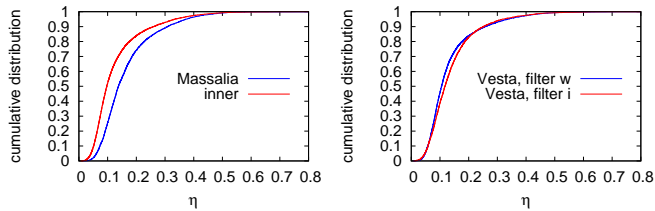}
\caption{Left: Cumulative distributions of brightness variation $\eta$ for Massalia family and its background. Right: The same for Vesta family in filter $w$ and filter $i$.}
\label{eta_mas_filters}
\end{figure}

Next, we compare distributions of dynamical families with their background. The family membership of asteroids was taken from \citet{2015aste.book..297N}. The background for a family is formed by asteroids from the same part of the main belt as the family (inner, middle, pristine, outer), which do not belong to any other family. We focused on 13 most populous families: Vesta, Massalia, Flora, Nysa Polana and Phocaea in the inner belt; Eunomia, Gefion and Maria in the middle belt; Koronis in the pristine belt; Themis, Eos, Hygiea and Alauda in the outer belt; see also Fig. \ref{5parts}. The typical number of asteroids (for which we have enough data) in a family is few thousands, for Vesta, Flora and Nysa Polana it is slightly more than ten thousands and for Phocaea and Alauda it is less than 1000 (the exact numbers are in Table \ref{ks_ps_n}). Unlike \citet{2016A&A...596A..57C}, who did not reveal any differences among families, we found that Massalia has a significantly different distribution of $b/a$ from its background, containing more elongated asteroids. Distributions are shown in Fig. \ref{massalia_phocaea} on the left. Significantly different are also cumulative distributions of brightness variation $\eta$ of Massalia and its background, which are shown in Fig. \ref{eta_mas_filters} on the left. Unfortunately, we cannot compare our distribution of $b/a$ for Massalia with the distribution from \citet{2017A&A...601A.139N} based on WISE data, because their sample contained insufficient number of bodies. The mean values of $D_{b/a}$ for all families are listed in Table~\ref{ks_ps}. The second largest difference between distribution of $b/a$ is for the Phocaea family and its background (see Fig. \ref{massalia_phocaea} on the right), nevertheless the value $D_{b/a}(L^1)=0.175$ is not high enough for a definite answer. We should note than for Phocaea we have only data for 812 asteroids, however, the small number of asteroids causes the population to be more spheroidal and, as we can see in Fig. \ref{massalia_phocaea} on the right, Phocaea, in comparison to its background, contains more elongated objects.

We have to remind that the difference between Massalia family and its background can be due to the different period distributions. To test this possibility we used the LCDB database and constructed distributions of $P$ for Massalia and its background. We found that Massalia really contains less objects with $P=0-4\,$h and more with $P=4-8\,$h than its background, which is in accordance with the family members being more elongated (compare with Fig. \ref{rot_period_real}). However, we have to emphasize that the distribution of $P$ for Massalia contains only 100 bodies and 420 bodies represent its background, which is not enough for a solid conclusion. For Phocaea family, we do not have enough determined periods to perform such test as for Massalia.

Our distributions of $b/a$ look different from the results of \citet{2008Icar..196..135S}, who determined distributions for eight families using data from the Sloan Digital Sky Survey (SDSS), however they assumed a fixed value of spin axis latitude for all asteroids, which probably influenced the results. We also did not find any dependence of the distribution of $b/a$ on the age of family that they suggested.

\begin{table}
 \small
\centering
\begin{tabular}{ccccc}
\hline
\hline
\rule{0cm}{2.5ex}
 family & $N_w$ & background$_w$ & $N_i$ & background$_i$ \\
\hline
\rule{0cm}{2.5ex}%
 Flora & 11 291 & 11 029 & 4135 & 5316\\
 Massalia & 4267 & 11 029 & 1032 & 5316\\
 Nysa Polana & 14 741 & 11 029 & 4675 & 5316\\
 Vesta & 11 895 & 11 029 & 4863 & 5316\\
 Phocaea & 812 & 11 029 & 577 & 5316\\
 Eunomia & 4126 & 12 069 & 2247 & 6728\\
 Gefion & 2629 & 12 069 & 1203 & 6728\\
 Maria & 2203 & 12 069 & 1243 & 6728\\
 Koronis & 4845 & 1272 & 1881 & 775\\
 Eos & 8237 & 6665 & 4272 & 4172\\
 Hygiea & 4191 & 6665 & 1584 & 4172\\
 Themis & 4181 & 6665 & 1588 & 4172\\
 Alauda & 649 & 6665 & 489 & 4172\\
\hline
\end{tabular}
\vspace{2mm}
\caption{\footnotesize{The number of asteroids in individual families and corresponding backgrounds for which we have data from Pan-STARRS1 survey in filters $w$ and $i$.}}
\label{ks_ps_n}
\end{table}

\subsection{Taxonomic classes and different parts of the main belt}
We also compared the distributions of $b/a$ of the two most populated taxonomic classes: S that dominates in the inner mail belt, and C that dominates in the middle and outer belt. We assigned a~taxonomic class to asteroids according to the SDSS-based Asteroid Taxonomy (Hasselmann et al. \citeyear{2010A&A...510A..43C}, data are available on Planetary Data System\footnote{https://sbn.psi.edu/pds/resource/sdsstax.html}). For both classes we had data for $\sim10\,000$ asteroids. We did not find these two groups to have different distributions of the shape elongation $b/a$.

Finally, we compared groups of asteroids with different semimajor axes (inner, middle, pristine, outer) and with different inclinations of orbit. None of the subpopulations is significantly different from others.

\subsection{Comparison of results from filters w and i}
\label{i_filter}
We also analyzed Pan-STARRS1 data in the $i$-filter ($\sim700-800\,$nm, Tonry et al. \citeyear{2012ApJ...750...99T}) and compared the results with the $w$-filter. We had data for 136 463 asteroids and on average, there were $\sim10$ measurements for one asteroid. We focused only on taxonomic classes and dynamical families. There were not enough asteroids to study the dependence of the elongation of asteroids on the diameter (only few asteroids were in the two subpopulations with the largest $D$).

\begin{figure}
 \centering
 \includegraphics[width=\hsize]{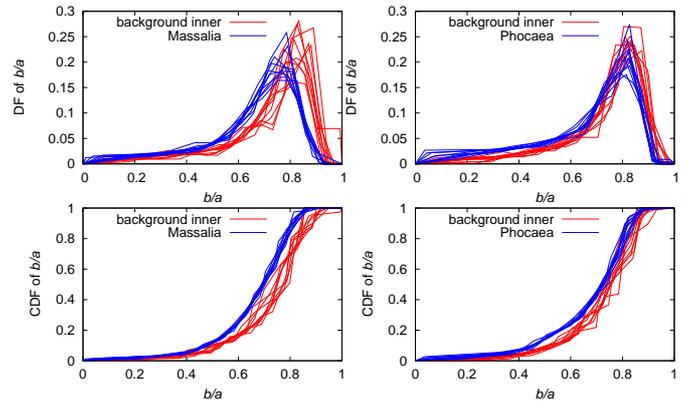}
 \caption{The same as in Fig. \ref{massalia_phocaea}, but in filter $i$.}
 \label{phocaea_i_filter}
\end{figure}

The number of asteroids in subpopulations containing the taxonomic class S was 6349 and for the taxonomic class C 5813. As in the $w$-filter, the difference between these two groups is insignificant. Then we focused on dynamical families. As in the $w$-filter, we found that Massalia family has a significantly different distribution of $b/a$ from its background. Moreover, also the result for Phocaea ($D_{b/a}(L^1)=0.212$) suggests that this family could have a different distribution of $b/a$ from its background. However, Fig.~\ref{phocaea_i_filter} does not show a significant difference.

\begin{figure}
\centering
\includegraphics[width=\hsize]{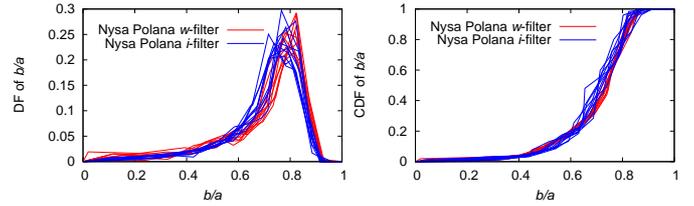}
\caption{DFs (left) and CDFs (right) of $b/a$ for Nysa Polana family in filter $i$ (blue lines) and filter $w$ (red lines).}
\label{w_vs_i}
\end{figure}

To compare results from the filters $w$ and $i$ directly, we constructed distributions of $b/a$ for some families in both filters and calculated $D_{b/a}$. We did not find any significant differences between filters. As an example, distributions of Nysa Polana are shown in Fig. \ref{w_vs_i}. We also constructed cumulative distributions of the brightness variation $\eta$ for some families in both filters to check that there are no differences between filters before the inversion (see Fig. \ref{eta_mas_filters} on the right).

\section{Conclusions}
In this work, we analyzed photometric data from the Pan-STARRS1 survey using a statistical approach based on cumulative distribution functions. We applied the model and the software package LEADER from \citet{2017A&A...601A.139N} and \citet{2017A&A..N} that allows us to construct distribution functions of the shape elongation $b/a$ and the ecliptical latitude $\beta$ of the spin axis for some subpopulations of asteroids and compare them with each other. Limitations of this model are: (i) it does not provide the pole longitude and (ii) it provides only the combined distribution of the $\beta$ of both ecliptic hemispheres. Moreover, by testing on synthetic data we found that our model shifts the peak of the $\beta$ distribution to the middle values and is strongly influenced by the number of objects in studied subpopulations. Distribution of $\beta$ also appears to be highly sensitive to the used database. For the distribution of $b/a$ we found that the model provides stable results for numbers of objects higher than $\sim 800$. The test with synthetic data also revealed that our model provides reliable results only for asteroids with rotation periods $P\lesssim 12\,$h. This is due to the time distribution of measurements of Pan-STARRS1 survey and thus it is not a limitation of the method in general.

We analyzed mainly data in the wide $w$-band filter. The most populous subpopulations were studied also in the $i$-filter. The main results of this paper are as follows:
\begin{enumerate}
 \item Groups of asteroids with diameter $D<25\,$km do not have significantly different distributions of $b/a$, the maximum of these distribution is for $b/a\simeq0.8$. The distribution for asteroids larger than 25 km suggests that these objects are more spheroidal in comparison with the smaller ones, nevertheless, the number of objects in this subpopulation is insufficient for a strong result.
 \item By comparing distributions of $b/a$ for different intervals of rotation period $P$ we found, that the fastest rotators with $P=0-4\,$h are more spheroidal (the maximum is for $b/a\sim 0.75$) than the population with $P=4-8\,$h (the maximum is for $b/a\sim 0.6$).
 \item We constructed distributions of $b/a$ for 13 most populous dynamical families. We revealed two families in the inner belt, Massalia and Phocaea, to be significantly different from their background. Both families have members that are more elongated than corresponding backgrounds. One possible explanation is that such result is due to the dependence of shape elongation on the rotation period.
 \item By analyzing data in the $i$-filter we confirmed previous results and we did not found any significant differences between subpopulations studied in the $w$-filter in comparison with the $i$-filter.
\end{enumerate}

\begin{acknowledgements}
      H. Cibulkov\'{a} and J. \v{D}urech were supported by the grant 15-04816S of the Czech Science Foundation. The research by H. Nortunen and M. Kaasalainen was supported by the Academy of Finland (Centre of Excellence in Inverse Problems), and H. N. was additionally supported by the grant of Jenny and Antti Wihuri Foundation. We would like to thank Matti Viikinkoski for valuable comments and feedback, as well as assistance with the software.
\end{acknowledgements}

%
 \bibliographystyle{aa} 
 \bibliography{bib_ps} 

\end{document}